\documentclass[twocolumn]{aastex63}

\received{}
\revised{}
\accepted{}

\submitjournal{AJ}

\shorttitle{Multiple populations in VLM stars}
\shortauthors{Dondoglio et al.}

\begin{document}

\title{Survey of multiple populations in globular clusters among very low-mass stars. }

\correspondingauthor{Emanuele Dondoglio}
\email{emanuele.dondoglio@phd.unipd.it}

\author[0000-0001-8415-8531]{E.\,Dondoglio} 
\affiliation{Dipartimento di Fisica e Astronomia ``Galileo Galilei'', Universit\`{a} di Padova, Vicolo dell'Osservatorio 3, I-35122, Padua, Italy}

\author[0000-0001-7506-930X]{A.\,P.\,Milone}                                    
\affiliation{Dipartimento di Fisica e Astronomia ``Galileo Galilei'', Universit\`{a} di Padova, Vicolo dell'Osservatorio 3, I-35122, Padua, Italy}             
\affiliation{Istituto Nazionale di Astrofisica - Osservatorio Astronomico di Padova, Vicolo dell'Osservatorio 5, IT-35122, Padua, Italy} 

\author[0000-0002-7093-7355]{A.\,Renzini}                                     
\affiliation{Istituto Nazionale di Astrofisica - Osservatorio Astronomico di Padova, Vicolo dell'Osservatorio 5, IT-35122, Padua, Italy}  
 
\author[0000-0003-2742-6872]{E\,Vesperini}                         
\affil{Department of Astronomy, Indiana University, Bloomington, IN 47401, USA}  

\author[0000-0003-1713-0082]{E.\,P.\,Lagioia}                                   
\affiliation{Dipartimento di Fisica e Astronomia ``Galileo Galilei'', Universit\`{a} di Padova, Vicolo dell'Osservatorio 3, I-35122, Padua, Italy} 

\author[0000-0002-1276-5487]{A.\,F.\,Marino}                                    
\affiliation{Istituto Nazionale di Astrofisica - Osservatorio Astronomico di Padova, Vicolo dell'Osservatorio 5, IT-35122, Padua, Italy} 

\author[0000-0003-3858-637X]{A.\,Bellini} \affil{Space Telescope Science Institute, 3700 San Martin Drive, Baltimore, MD 21218, USA}  

\author[0000-0003-1757-6666]{M.\,Carlos}                                                 
\affiliation{Dipartimento di Fisica e Astronomia ``Galileo Galilei'', Universit\`{a} di Padova, Vicolo dell'Osservatorio 3, I-35122, Padua, Italy} 

\author[0000-0002-7690-7683]{G.\,Cordoni}                                                
\affiliation{Dipartimento di Fisica e Astronomia ``Galileo Galilei'', Universit\`{a} di Padova, Vicolo dell'Osservatorio 3, I-35122, Padua, Italy}  

\author[0000-0002-1562-7557]{S.\,Jang}                                                   
\affiliation{Dipartimento di Fisica e Astronomia ``Galileo Galilei'', Universit\`{a} di Padova, Vicolo dell'Osservatorio 3, I-35122, Padua, Italy}  

\author[0000-0003-3153-1499]{M.\,V.\,Legnardi}                                                              
\affiliation{Dipartimento di Fisica e Astronomia ``Galileo Galilei'', Universit\`{a} di Padova, Vicolo dell'Osservatorio 3, I-35122, Padua, Italy}

\author[0000-0001-9673-7397]{M\,Libralato} \affil{AURA for the European Space Agency (ESA), ESA Office, Space Telescope Science Institute, 3700 San Martin Drive, Baltimore, MD21218, USA}   

\author[0000-0001-5182-0330]{A.\,Mohandasan}                                                                  
\affiliation{Dipartimento di Fisica e Astronomia ``Galileo Galilei'', Universit\`{a} di Padova, Vicolo dell'Osservatorio 3, I-35122, Padua, Italy}

\author[0000-0003-4697-0945]{F. D'Antona}                                       
\affil{INAF - Osservatorio Astronomico di Roma, Via Frascati 33, I-00040, Monte Porzio Catone, Roma, Italy}

\author[0000-0003-2373-0404]{M.\,Martorano}                                              
\affiliation{Dipartimento di Fisica e Astronomia ``Galileo Galilei'', Universit\`{a} di Padova, Vicolo dell'Osservatorio 3, I-35122, Padua, Italy}

\author{F.\,Muratore}                                              
\affiliation{Dipartimento di Fisica e Astronomia ``Galileo Galilei'', Universit\`{a} di Padova, Vicolo dell'Osservatorio 3, I-35122, Padua, Italy}

\author[0000-0002-1128-098X]{M.\,Tailo}                                                  
\affiliation{Dipartimento di Fisica e Astronomia ``Galileo Galilei'', Universit\`{a} di Padova, Vicolo dell'Osservatorio 3, I-35122, Padua, Italy}

\begin{abstract}
Recent work has shown that NIR {\it Hubble Space Telescope} ({\it HST}) photometry allows us to disentangle multiple populations (MPs) among M dwarfs of globular clusters (GCs) and investigate this phenomenon in very low-mass (VLM) stars. Here, we present the color-magnitude diagrams (CMDs) of nine GCs and the open cluster NGC\,6791 in the F110W and F160W bands of HST, showing that the main sequences (MSs) below the knee are either broadened or split thus providing evidence of MPs among VLM stars. 
In contrast, the MS of NGC\,6791 is consistent with a single population. The color distribution of M-dwarfs dramatically changes between different GCs and the color width correlates with the cluster mass. We conclude that the MP ubiquity, variety, and dependence on GC mass are properties common to VLM and more-massive stars.

We combined UV, optical, and NIR observations of NGC\,2808 and NGC\,6121 (M\,4) to identify MPs along with a wide range of stellar masses ($\sim0.2-0.8 \mathcal{M}_{\odot}$), from the MS turn off to the VLM regime, and measured, for the first time, their mass functions (MFs). We find that the fraction of MPs does not depend on the stellar mass and that their MFs have similar slopes. These findings indicate that the properties of MPs do not depend on stellar mass. In a scenario where the second generations formed in higher-density environments than the first generations, the possibility that the MPs formed with the same initial MF would suggest that it does not depend on the environment.
\end{abstract}


\section{Introduction} \label{sec:intro}
It is widely accepted that globular-cluster (GC) stars can be classified into first-generation (1G) stars, with halo-like chemical composition, and second-generation (2G) stars characterized by distinctive light-element abundances. 
The 2G is composed of stars depleted in some light elements, including carbon and oxygen and enriched in helium, nitrogen and sodium with respect to the 1G \citep[e.g.,][]{kraft1994a, gratton2004a, carretta2009a, milone2017b, marino2019a}.

The origin of multiple stellar populations (MPs) is still an open question. To explain the chemical composition of 2G stars two main groups of formation scenarios have been proposed so far.
The first one foresees multiple star formation episodes: throughout their evolutionary path, the intermediate- to high- mass stars that formed during the first burst eject winds of processed material out of which 2G stars form (with the possible contribution of gas with pristine composition).
Many polluters have been proposed including intermediate-mass asymptotic giant branch (AGB) stars, fast-rotating massive stars, supermassive stars or massive interacting binaries \citep[e.g.,][]{ventura2001a, dantona2016a, demink2009, decressin2007a, dercole2010a, krause2013, denissenkov2014, calura2019a}.

The second category involves early accretion of material ejected by supermassive stars \citep[][]{denissenkov2014} on forming protostars \citep[see e.g.,][]{gieles2018} and suggests that all the stars form in a single episode. 
In addition, stellar mergers have been recently proposed as possible responsible for MPs in GCs \citep[][]{wang2020a}.

The poorly-explored very-low mass (VLM) stars regime would provide crucial insights to discriminate among the formation scenarios. VLM stars have masses smaller than $\sim 0.4 \mathcal{M}_{\rm \odot}$, and are characterized by a high density and a low effective temperature, with their spectral peak in the near-infrared (NIR), where various molecules including oxygen (e.g.\, CO, $\rm{H_{\rm 2}O}$, OH, TiO, VO, ZrO) are primary sources of opacity \citep[e.g.][]{allard1995}. Being among the faintest stars that one can detect in GCs, their observation is particularly challenging. For this reason, while MPs have been widely studied among stars more massive than $\sim 0.6 \mathcal{M}_{\rm \odot}$, the M-dwarfs regime is almost unexplored. To date, indeed, MPs have been identified and chemically characterized through photometric studies only in four clusters, namely NGC\,2808, NGC\,6121 (M\,4) and NGC\,5139 ($\omega$\,Cen) and NGC\,6752 \citep[][]{milone2012a, milone2014, milone2017a, dotter2015a, bellini2018, milone2019a}.
This was possible by exploiting the NIR camera on board {\it Hubble Space Telescope} ({\it HST}), thanks to the $m_{\rm F110W}-m_{\rm F160W}$ color, which is sensitive to molecular absorption bands (in particular to $\rm{H_{\rm 2}O}$) and therefore is effective to disentangle the oxigen-different 1G and 2G stars.

Separating MPs in VLM stars is also crucial to extend the study of their mass functions (MFs) to such low stellar masses and provide solid slope estimates. To date, the only observational study on MF of MPs has been carried out by \citet{milone2012b} for three populations with different helium in NGC\,2808. They have not found any significant differences between the MF slopes of the distinct populations, although the stars with pristine helium abundances seemed to deviate from a power law, flattening below $\sim 0.6 \mathcal{M}_{\rm \odot}$. The dataset used in this first pioneering work covered a narrow stellar mass range ($\sim 0.75-0.45 \mathcal{M}_{\rm \odot}$), preventing Milone and collaborators from drawing strong conclusions. Extending the study through lower stellar masses is instrumental in deriving firm conclusions on the different-population MFs.

At the same time, determining the MF slopes of MPs can provide invaluable insights into the differences in their formation and dynamical history. A careful interpretation of the present-day MF of MPs requires taking into account possible differences induced by dynamical processes  (mass loss and mass segregation) on the global and local (i.e. measured at a given clustercentric distance) MF  along with those which might instead due to possible differences in the initial mass function (IMF) and arose at the time of the cluster formation. \citet{vesperini2018}, by means of N-body simulations, have studied the evolution of the MF in clusters with MPs and explored the extent of expected variations in the MF arising from the effects of dynamics in different stellar populations starting with the same IMF and those which, instead, require MPs to form with different IMFs.

The comparison of the MP properties over a wide mass range provides several key constraints on the formation scenarios. Indeed, in scenarios based on the accretion of processed gas on protostars, the 2G chemical composition would depend on the mass of the protostellar object: by assuming a Bondy-Hoyle-Littleton accretion, the amount of accreted material is proportional to the square of the stellar mass. Consequently, less massive stars would accrete a smaller amount of processed material, and exhibit smaller internal variation of light elements than massive stars. Finally, a stellar population forming from Bondi accretion is expected to follow a MF with a slope equal to -2 \citep[e.g.,][]{ballesteros2015} which, in the low-mass regime, is significantly different from the slope (-1.3) of a \citet[][]{2001kroupa} IMF.

Driven by these results, we start investigating deep NIR {\it{HST}} observations of nine Galactic GCs and one Galactic open cluster to explore their VLM stars and perform an early census of MPs in this mass regime. 
To our knowledge, this sample comprises all clusters for which either proprietary or public appropriate NIR data are available in the {\it HST} archive. 
In this paper, we describe the dataset and present the NIR CMDs. Moreover, we derive the MF of different stellar populations in the GCs NGC\,2808 and M4, where MPs among M-dwarfs have been detected and chemically characterized in previous works \citep[][]{milone2012b, milone2014}.
The paper is organized as follows: Section~\ref{sec:data} describes the dataset and summarizes the procedure for data reduction, the NIR CMDs of all the clusters are presented in Section~\ref{sec:panel}. In Section~\ref{sec:selection} we present the two GCs where we measured the MP MFs, NGC\,2808 and M4, introducing the photometric tools that we exploited to separate MPs along the main sequence (MS). Section~\ref{sec:lf} and Section~\ref{sec:lfM4} describe how we derived the MFs of stellar populations in NGC\,2808 and M\,4, respectively. Section~\ref{sec:rad} explores the radial behaviour of MP pattern in both GCs and Section~\ref{sec:res} discusses and summarizes the results.

\begin{table*}[ht]
\caption{Summary of the data used in this work. The table lists, for each cluster, the average NIR FoV coordinate (J2000) and distance from cluster centre (in arcmin), the exposure times, filters and cameras used for each image, and the program.}

\scriptsize
\begin{tabular}{cccccc}
\label{dataset}
\\
\hline\hline
     CLUSTER   & Distance & N$\times$ EXPTIME & FILTER & INSTRUMENT & PROGRAM \\
     (RA, Dec) & [arcmin] &                   &        &            &         \\
    \hline
         \\
     NGC\,104                                       &  5.98 & 18$\times$149s & F110W & IR/WFC3 & 11443 \\
     (00h:22m:29.60s, $-$72$^{\circ}$:04':05.02'')  & & 499s           & F110W & IR/WFC3 & 11926 \\
                                                    & & 42$\times$274s & F160W & IR/WFC3 & 11443-5 \\
                                                    & & 24$\times$92s+24$\times$352s & F160W & IR/WFC3 & 11931 \\ 
                                                    & & 14$\times$92s+6$\times$352s  & F160W & IR/WFC3 & 12352 \\ 
                                                    & & 14$\times$92s+6$\times$352s  & F160W & IR/WFC3 & 12696 \\ 
                                                    & & 4$\times$92s+2$\times$352s   & F160W & IR/WFC3 & 13079 \\ 
                                                    & & 4$\times$92s+2$\times$352s   & F160W & IR/WFC3 & 13563 \\ 

     \\
     NGC\,288                                       & 5.76 & 15s+3$\times$200s & F606W & WFC/ACS & 12193 \\
     (00h:52m:22.75s, $-$26$^{\circ}$:36':52.78'')  & & 10s+3$\times$150s & F814W & WFC/ACS & 12193\\
                                                    & & 3$\times$142s+5$\times$1202s & F110W & IR/WFC3 & 16289\\
                                                    & & 4$\times$142s+2$\times$1202+7$\times$1302s & F160W & IR/WFC3 & 16289 \\
     \\
     NGC\,1851                                      & 3.11 & 2$\times$357s              & F606W & WFC/ACS & 10458 \\
     (05h:13m:52.92s, $-$40$^{\circ}$:04':27.61'')  & & 2$\times$32s+3$\times$899s & F110W & IR/WFC3 & 16177 \\
                                                    & & 2$\times$32s+99s+3$\times$1599s & F160W & IR/WFC3 & 16177\\

     \\
     NGC\,2808 - Field A                            & 5.31 & 4$\times$50s+2$\times$620s+2$\times$655s & F390W & UVIS/WFC3 & 11665 \\
     (09h:11m:21.48s, $-$64$^{\circ}$:54':48.01'')  & & 2$\times$699s & F110W & IR/WFC3 & 11665\\
                                                    & & 799s+899s & F160W & IR/WFC3 & 11665 \\
     NGC\,2808 - Field B                            & 5.21 & 4$\times$50s+2$\times$620s+2$\times$655s & F390W & UVIS/WFC3 & 11665 \\
     (09h:11m:56.67s, $-$64$^{\circ}$:56':56.29'')  & & 2$\times$699s & F110W & IR/WFC3 & 11665 \\
                                                    & & 799s+899s & F160W & IR/WFC3 & 11665 \\
     NGC\,2808 - Field C                            & 5.48 & 4$\times$50s+2$\times$620s+2$\times$655s & F390W & UVIS/WFC3 & 11665\\
     (09h:12m:49.87s, $-$64$^{\circ}$:49':36.02'')  & & 2$\times$699s & F110W & IR/WFC3 & 11665\\
                                                    & & 799s+899s & F160W & IR/WFC3 & 11665\\
     \\
     NGC\,5139                                      & 16.72 & 2$\times$1300s+2$\times$1375s & F606W & WFC/ACS &  9444 \\
     (13h:25m:36.61s, $-$47$^{\circ}$:39':50.38'')  & & 2$\times$1340s+2$\times$1375s & F814W & WFC/ACS &  9444 \\
                                                    & & 2$\times$1285s+2$\times$1331s & F606W & WFC/ACS & 10101 \\
                                                    & & 4$\times$1331s                & F814W & WFC/ACS & 10101 \\
                                                    & & 7$\times$142s+14$\times$1302s & F110W & IR/WFC3 & 14118 \\
                                                    & & 7$\times$142s+14$\times$1302s & F160W & IR/WFC3 & 14118 \\
     \\
     NGC\,5904                                      & 5.70 & 621s                         & F475W & WFC/ACS & 13297 \\
     (15h:18m:56.03s,  02$^{\circ}$:03':49.42'')    & & 559s                         & F814W & WFC/ACS & 13297 \\
                                                    & & 2$\times$122s+4$\times$1202s & F110W & IR/WFC3 & 16289 \\
                                                    & & 3$\times$122s+6$\times$1302s & F160W & IR/WFC3 & 16289 \\
     \\
     NGC\,6121                                      & 1.94 & 4$\times$680s & F275W & UVIS/WFC3 & 16289 \\
     (16h:23m:41.57s, $-$26$^{\circ}$:30':29.43'')  & & 4$\times$358s & F336W & UVIS/WFC3 & 16289 \\
                                                    & & 4$\times$105s & F438W & UVIS/WFC3 & 16289 \\
                                                    & & 8$\times$652s & F110W & IR/WFC3 & 12602 \\
                                                    & & 8$\times$652s & F110W & IR/WFC3 & 14752 \\
                                                    & & 16$\times$652s & F160W & IR/WFC3 & 12602 \\
     \\
     NGC\,6656                                      & 6.19 & 2$\times$656s & F475W & ACS/WFC & 12311 \\
     (18h:36m:45.00s, $-$23$^{\circ}$:58':10.02'')  & & 2$\times$389s & F814W & ACS/WFC & 12311 \\
                                                    & & 32s+3$\times$124s+149s+249s                    & F110W & IR/WFC3 & 16177 \\
                                                    & & 2$\times$149s+3$\times$174s+199s+2$\times$249s & F160W & IR/WFC3 & 16177 \\
     \\
     NGC\,6752                                      & 4.89 & 28$\times$142s+56$\times$1302s & F110W & IR/WFC3 & 15096 \\
     (19h:11m:19.66s, $-$59$^{\circ}$:55':37.45'')  & & 17$\times$142s+34$\times$1302s & F160W & IR/WFC3 & 15096 \\
                                                    & & 28$\times$142s+56$\times$1302s & F110W & IR/WFC3 & 15491 \\
                                                    & & 12$\times$142s+24$\times$1302s & F160W & IR/WFC3 & 15491 \\
     \\
     NGC\,6791                                      & 0.59 & 3$\times$39s$+$2$\times$1142s$+$3$\times$1185s   & F606W & ACS/WFC &  9815 \\
     (19h:20m:53.95s,  37$^{\circ}$:48':09.60'')    & & 3$\times$39s$+$2$\times$1142s$+$3$\times$1185s   & F814W & ACS/WFC &  9815 \\
                                                    & & 3$\times$49s$+$260s$+$2$\times$399s   & F110W & IR/WFC3 & 11664 \\
                                                    & & 3$\times$49s$+$260s$+$2$\times$399s   & F160W & IR/WFC3 & 11664 \\
     \\
\hline  \hline
\end{tabular}
\label{tab:img}
\end{table*}

\section{Data and Data Reduction} \label{sec:data}

To investigate stellar populations among VLM stars of the ten studied clusters we used images collected through the F110W and F160W bands of the Near Infrared Channel of the Wide Field Camera 3 (NIR/WFC3) on board {\it{HST}}. 
We also used optical images collected through the Wide Field Channel of the Advanced Camera for Survey (WFC/ACS) to derive stellar proper motions and separate cluster members from the bulk of field stars in NGC\,104, NGC\,288, NGC\,1851, $\omega$\,Cen, NGC\,5904, M\,4,  NGC\,6656, NGC\,6752 and NGC\,6791. 
In addition, we exploited the Ultraviolet and Visual (UVIS) channel of WFC3 to disentangle multiple stellar populations in the upper MSs of NGC\,2808 and M\,4. 

To optimize the photometry of VLM stars, we analyzed fields of views (FoVs) that are far away from the cluster centers. The radial distance ranges from $\sim0.6$ arcmin for NGC\,6791 to $\sim16.7$ arcmin in $\omega$\,Cen. In the case of NGC\,2808 we analyzed three FoVs, namely A, B and C, located southwest, south, and northeast the cluster center, respectively.  
The main information about the dataset is provided in Table~\ref{dataset}.

We derived stellar positions and magnitudes by performing effective Point Spread Function (PSF) photometry \citep[e.g.][]{Anderson2000a} by means of the KS2 computer program, which is the evolved version of $\tt{kitchen_{-}sync}$, developed by Jay Anderson \citep[][]{anderson2008a}. 
KS2 adopts distinct methods to measure stars. In the method I, which provides the optimal photometry and astrometry of bright stars, each star is measured by using the PSF model corresponding to its position.  Stellar fluxes and positions are 
 derived in each exposure independently, and are then averaged together to derive the best estimates of magnitudes and coordinates.
Methods II and III combine information from all images together and provide robust measurements for faint stars. After subtracting neighbor stars, by using the most accurate estimate of stellar positions and fluxes, these methods perform aperture photometry over a region of either 5$\times$5 pixels (method II) or 0.75$\times$0.75 pixels (method III). The aperture center corresponds to the best determination of the stellar position. Clearly, method III works better in crowded environments \citep[see e.g.,][for detailed discussion]{sabbi2016a, bellini2017a, nardiello2018}.   

We calibrated the WFC3 magnitudes into the Vega system as in \citet{bedin2005}, by using the zero points available in the Space Telescope Science Institute webpage\footnote{\url{https://www.stsci.edu/hst/instrumentation/wfc3/data-analysis/photometric-calibration}} for UVIS/WFC3 and NIR/WFC3.

To identify and characterize the MPs in the CMDs, we selected stars that are well measured. To do this, we exploited the various diagnostics of the photometric and astrometric quality provided by the KS2. Specifically, we adopted the random mean scatter of the photometric measurements, the QFIT parameter, which indicates whether a star is well reproduced by a PSF model or not, \citep[e.g.][]{anderson2008a, nardiello2018} and the RADXS parameter \citep[][]{bedin2008}, which is an excellent tool to disentangle stellar sources from sources with broad (e.g. galaxies) or narrow profiles (e.g. cosmic rays and artifacts). We emphasize that these parameters, which are derived from the comparison of the observed sources with the PSF model, have been calculated for all methods separately. In method I we used the stellar fluxes derived from PSF fitting, while in methods II and III, where stellar fluxes are measured by means of aperture photometry, we adopted for the PSF model the magnitudes inferred with these methods. We refer to the paper by \citet[][see their Section 2 and references therein]{nardiello2018} for details.

\begin{figure*}[htp]
\centering
\includegraphics[height=11cm,trim={0.7cm 5cm 5.7cm 4cm},clip]{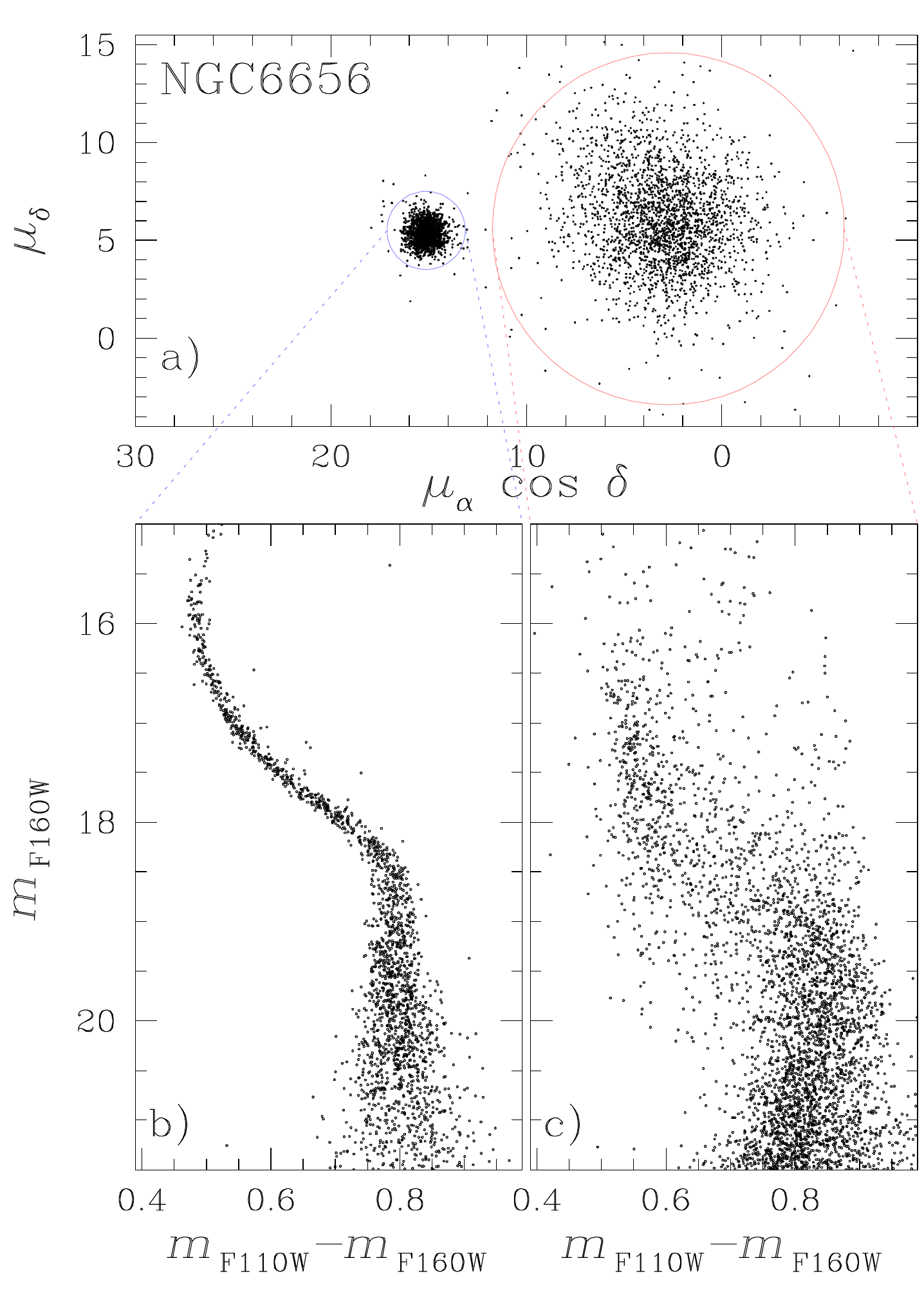}
\includegraphics[height=11cm,trim={0.7cm 5cm 6.7cm 4cm},clip]{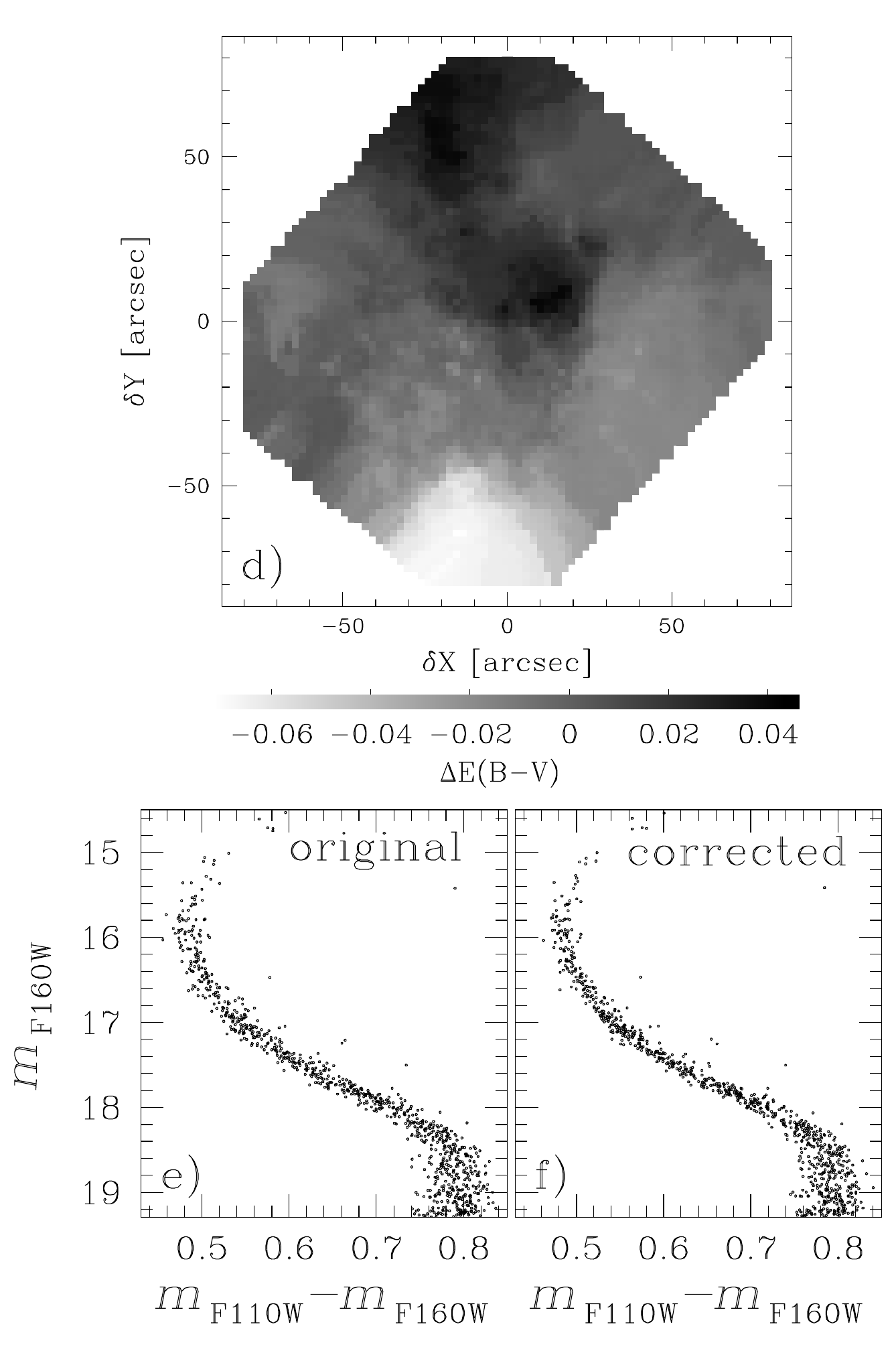}
\caption{This figure illustrates various steps for the determination of differential-reddening corrected photometry of cluster members in NGC\,6656. Panel a shows the vector-point diagram of proper motions (in mas/yr) for all stars in the FoV, while panels b and c show the $m_{\rm F160W}$ vs.\,$m_{\rm F110W}-m_{\rm F160W}$ CMDs for proper-motion selected cluster members and field stars, respectively. The map of differential reddening is plotted in  panel d, where the levels of gray are proportional to the amount of E(B$-$V) variation as indicated on the bottom. The comparison between the original CMD and the CMD corrected for differential reddening is provided in panels e and f, respectively.}
\label{fig:DRn6656}
\end{figure*}

The proper motions are measured by using the procedure and the computer programs by \citet[][]{piotto2012a} and are used to separate probable cluster members from field stars.  We averaged together the coordinates from all exposures of each epoch and compared the stellar positions in the different epochs  to infer the displacements relative to the bulk of cluster stars.  To transform proper motions from relative into absolute we considered stars for which both {\it HST}-based relative proper motions and absolute proper motions from Gaia eDR3 are available. We calculate the median difference between relative and absolute motions and added these quantities to the relative proper motion of each star.
As an example, we provide in Figure \ref{fig:DRn6656} the vector-point diagram of proper motions for stars in the FoV of NGC\,6656 (panel a), the $m_{\rm F160W}$ vs.\,$m_{\rm F110W}-m_{\rm F160W}$ CMDs of probable cluster members (panel b) and field stars (c).

Finally, we corrected our photometry for the effects of differential reddening and zero point spacial variations following the recipe by \citet[][see their Sections 3.1 and 3.2]{milone2012c}. In a nutshell, we derived the MS fiducial line and measured the displacement of bright MS stars from the fiducial along the reddening direction. The best differential-reddening estimate associated to each star, corresponds to the median displacement of its 50 nearest bright MS stars.
 Panel d in Figure \ref{fig:DRn6656} shows the differential-reddening map in the field of view of NGC\,6656, which is the studied GC with the largest reddening variation. Panels e and f compare the CMDs of the upper MS, which is the region where the effects of differential reddening are more evident, before and after the differential-reddening correction.  
 
\subsection{Artificial Star tests} \label{subsec:as}

Artificial-Star (AS) tests have been performed by following the recipe by \citet{anderson2008a} and used to estimate the photometric errors of all clusters and the completeness level of the photometry in NGC\,2808 and M\,4. 

In a nutshell, AS tests consist in adding into the images artificial stars with known positions and magnitudes and measure them by using the same procedure adopted for real stars. The measured magnitude and position of each AS are then compared with the input ones to evaluate whether the procedure has found that star and estimate the accuracy of the photometry and astrometry.
 
To perform AS test, we generated a catalog of 50000 stars with fixed positions and magnitudes. Stars are randomly distributed within each FoV with the criterion of mimicking the radial distribution of the observed stars. The magnitude of ASs are derived so that stars are placed on the fiducial lines that reproduce the average distribution of cluster stars in the observed CMDs.

A star is considered recovered if the distance between the input and output position and magnitude is less than 0.5 pixel and 0.75 magnitudes, respectively, and if it passes the criteria of selection adopted for real stars.
Completeness is calculated as the fraction of recovered stars on the input stars in different F160W magnitude bins.

\section{Near-Infrared Color-Magnitude Diagrams} \label{sec:panel}
\begin{figure*}[h]
\centering
\includegraphics[width=8.7cm,trim={0.7cm 6.7cm 0.2cm 11.9cm},clip]{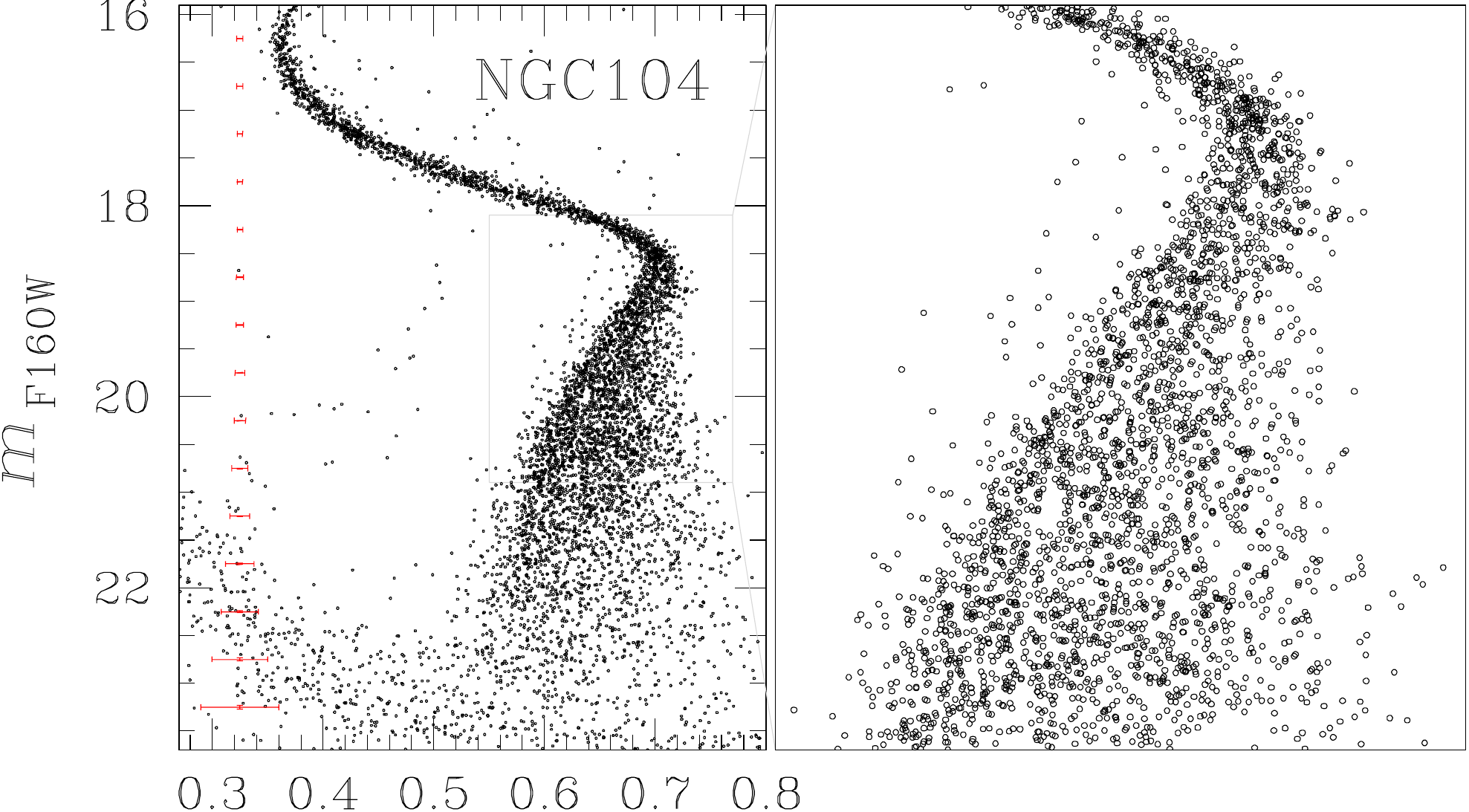}
\includegraphics[width=8.7cm,trim={0.7cm 6.7cm 0.2cm 11.9cm},clip]{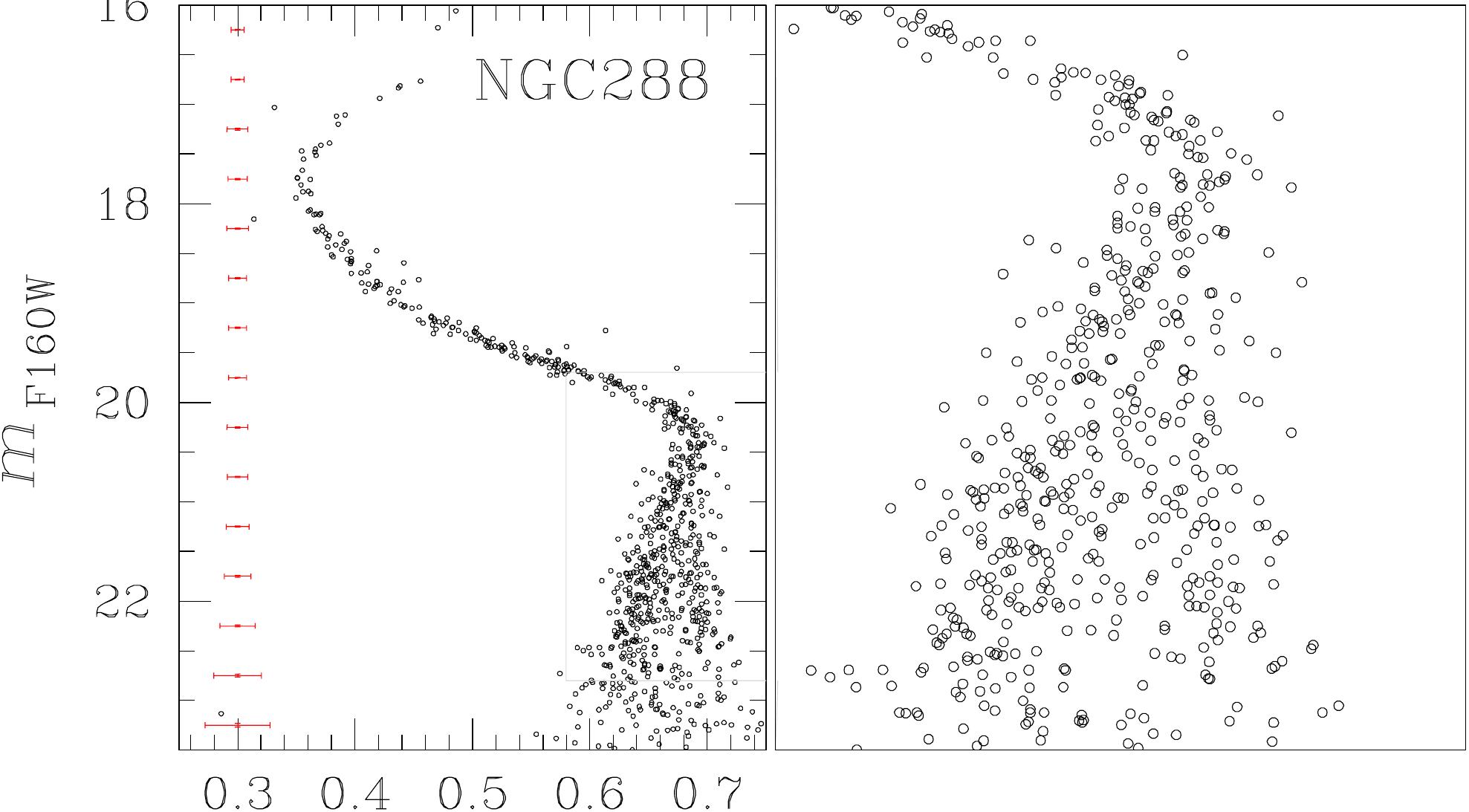}
\includegraphics[width=8.7cm,trim={0.7cm 6.7cm 0.2cm 11.9cm},clip]{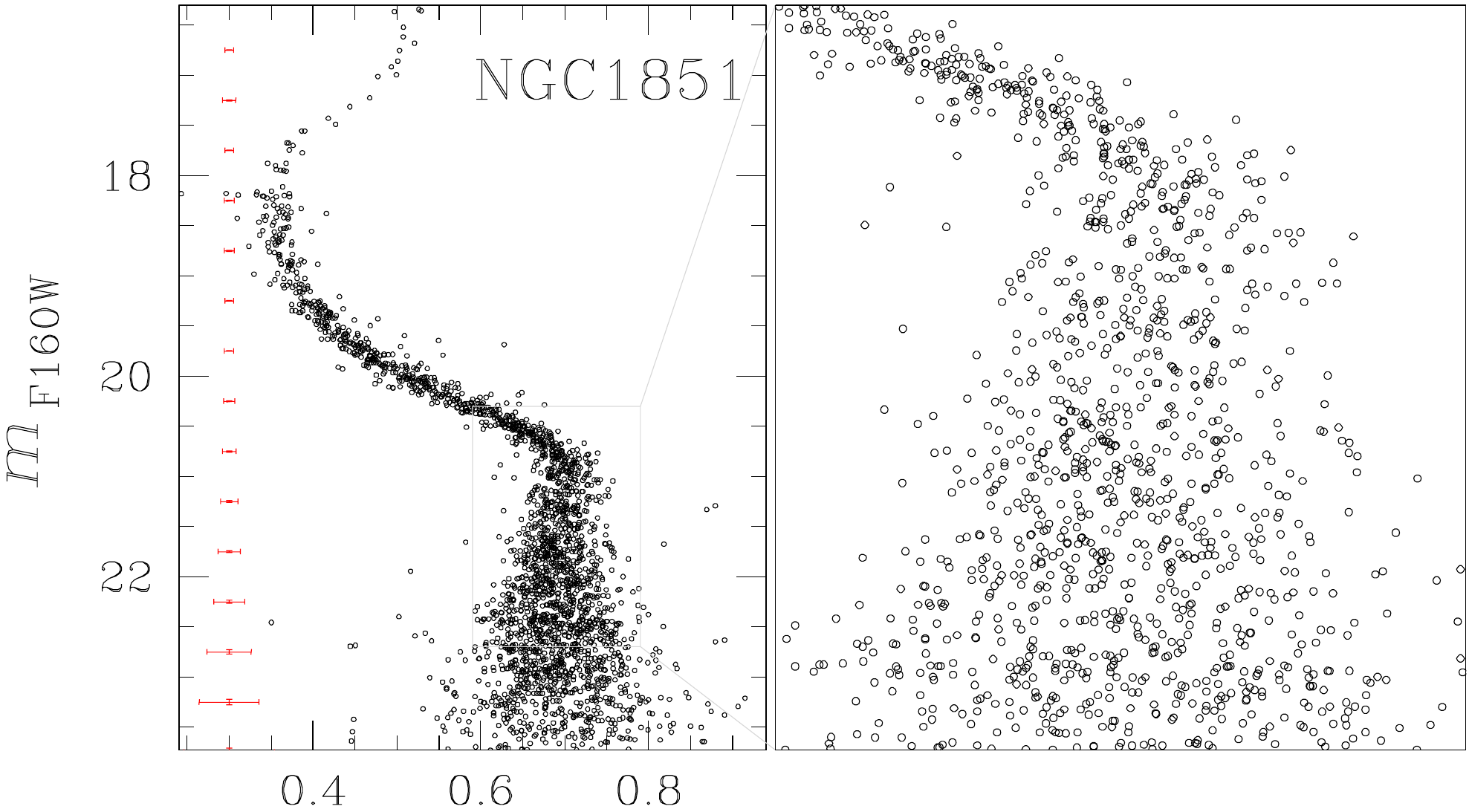}
\includegraphics[width=8.7cm,trim={0.7cm 6.7cm 0.2cm 11.9cm},clip]{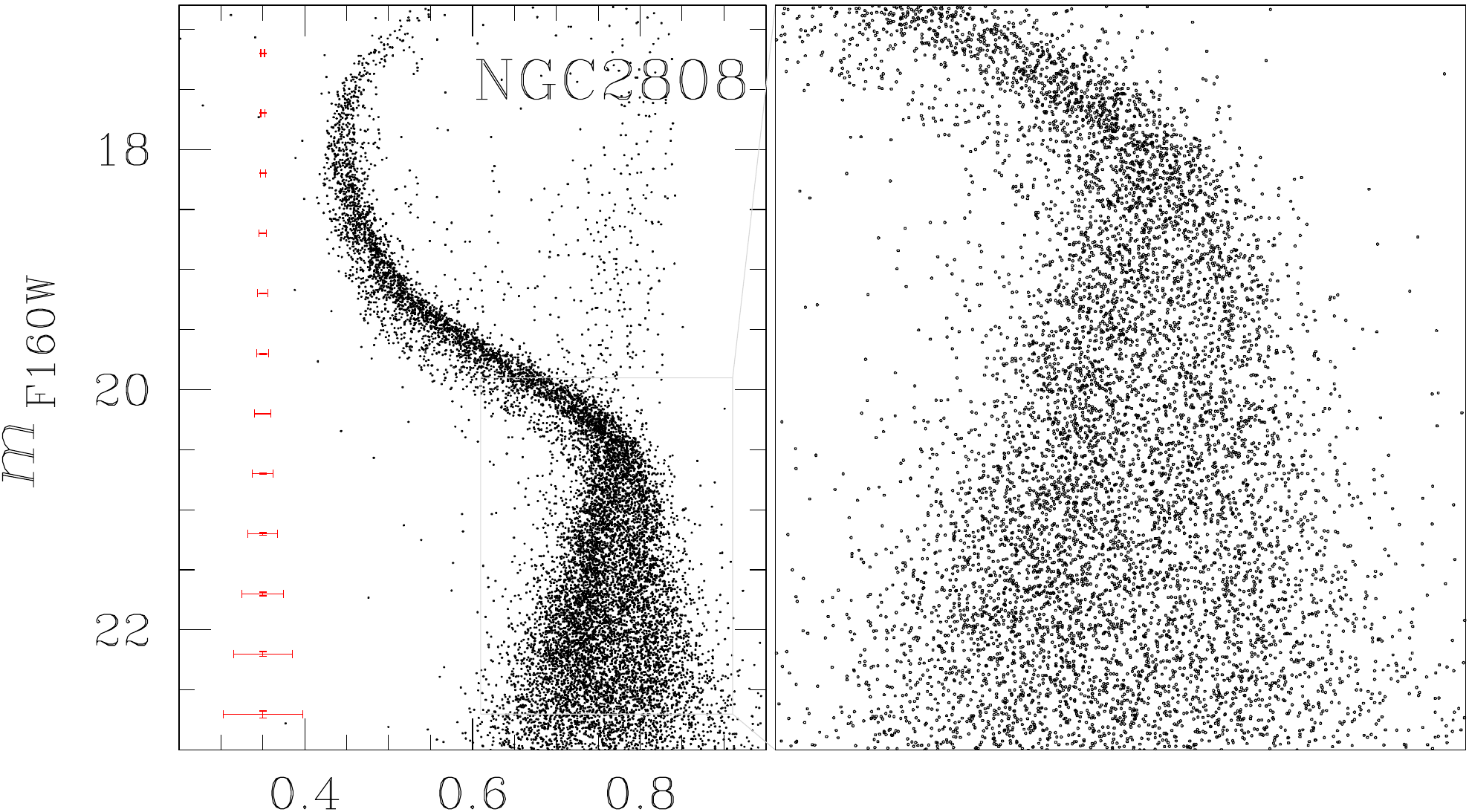}
\includegraphics[width=8.7cm,trim={0.7cm 6.7cm 0.2cm 11.9cm},clip]{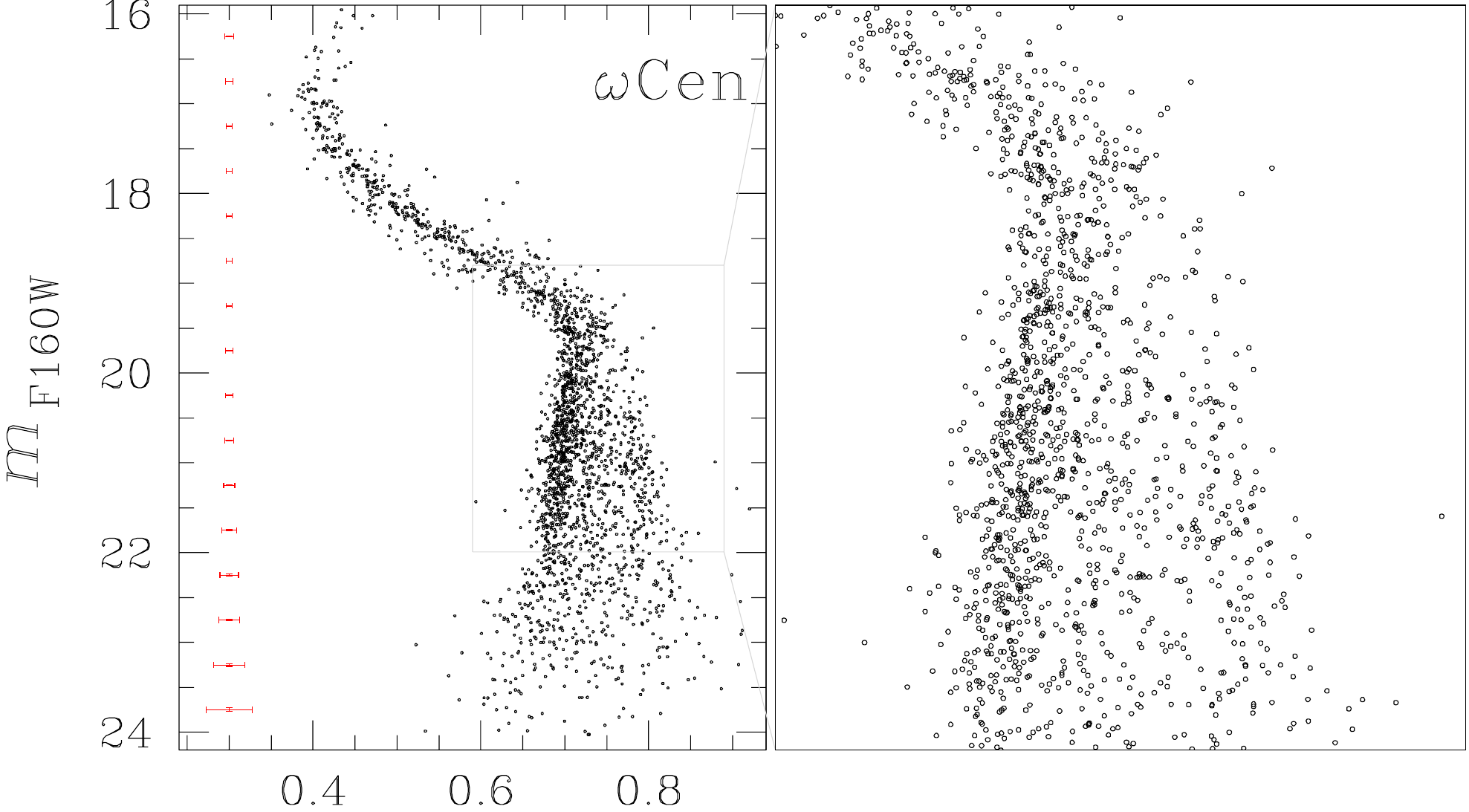}
\includegraphics[width=8.7cm,trim={0.7cm 6.7cm 0.2cm 11.9cm},clip]{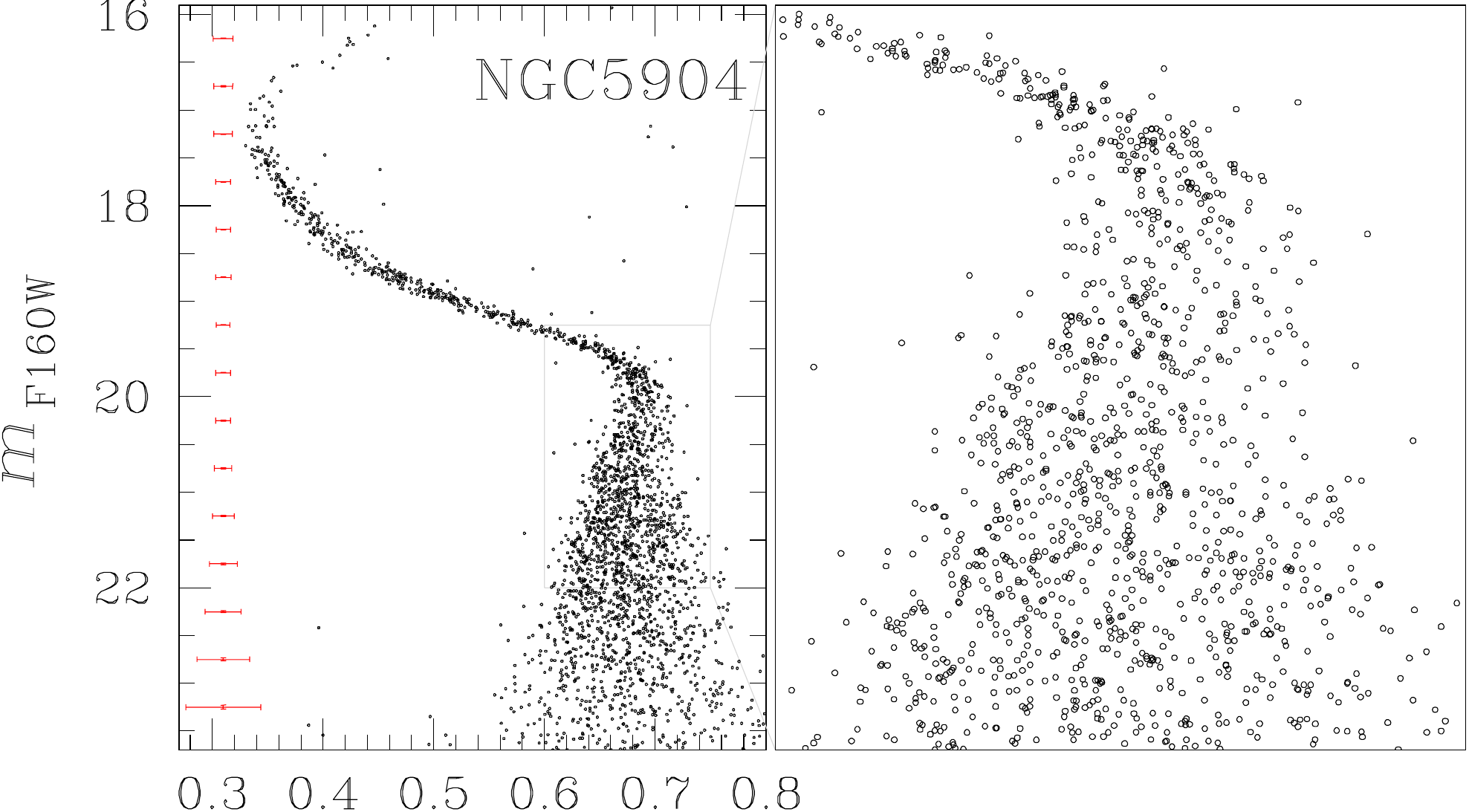}
\includegraphics[width=8.7cm,trim={0.7cm 6.7cm 0.2cm 11.9cm},clip]{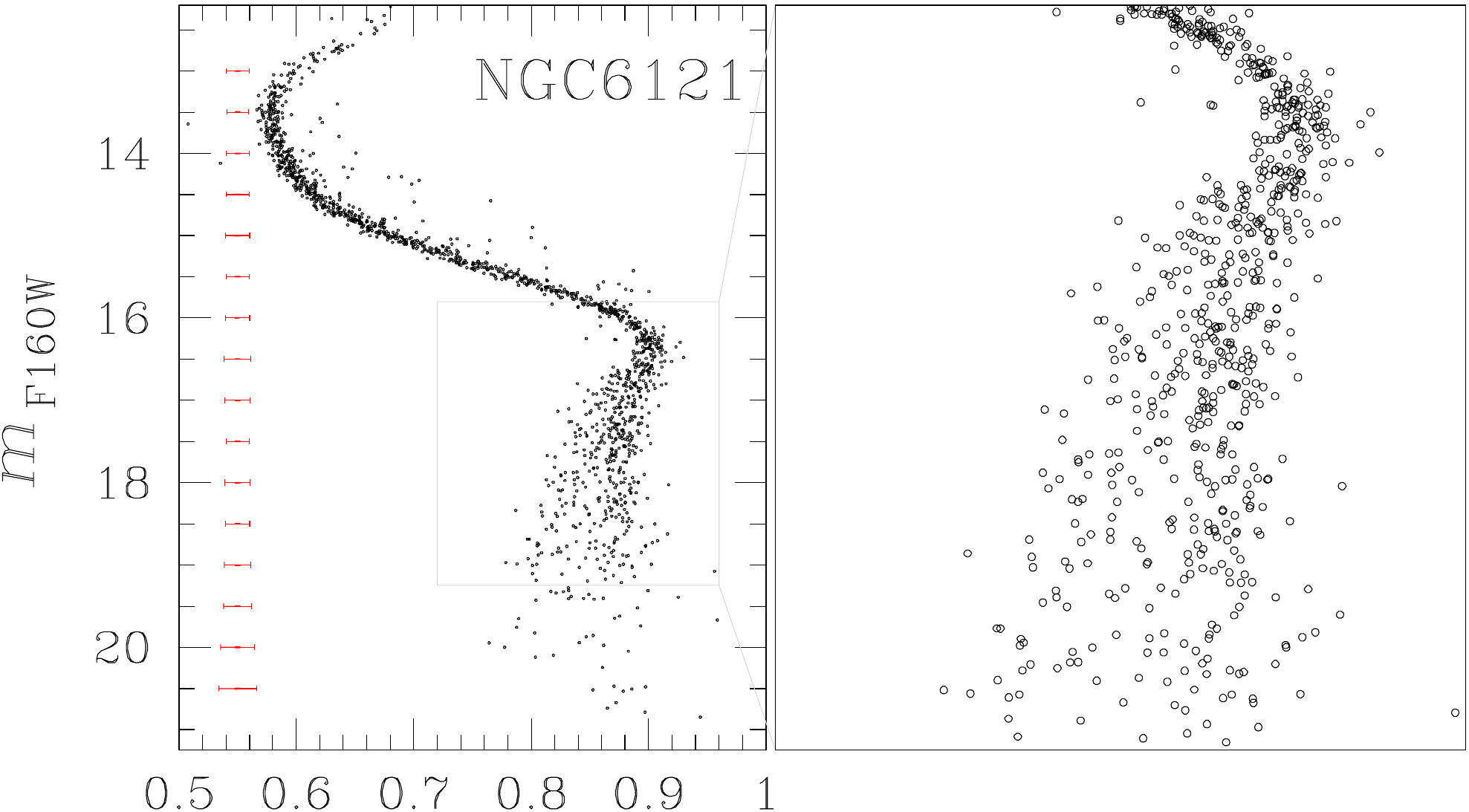}
\includegraphics[width=8.7cm,trim={0.7cm 6.7cm 0.2cm 11.9cm},clip]{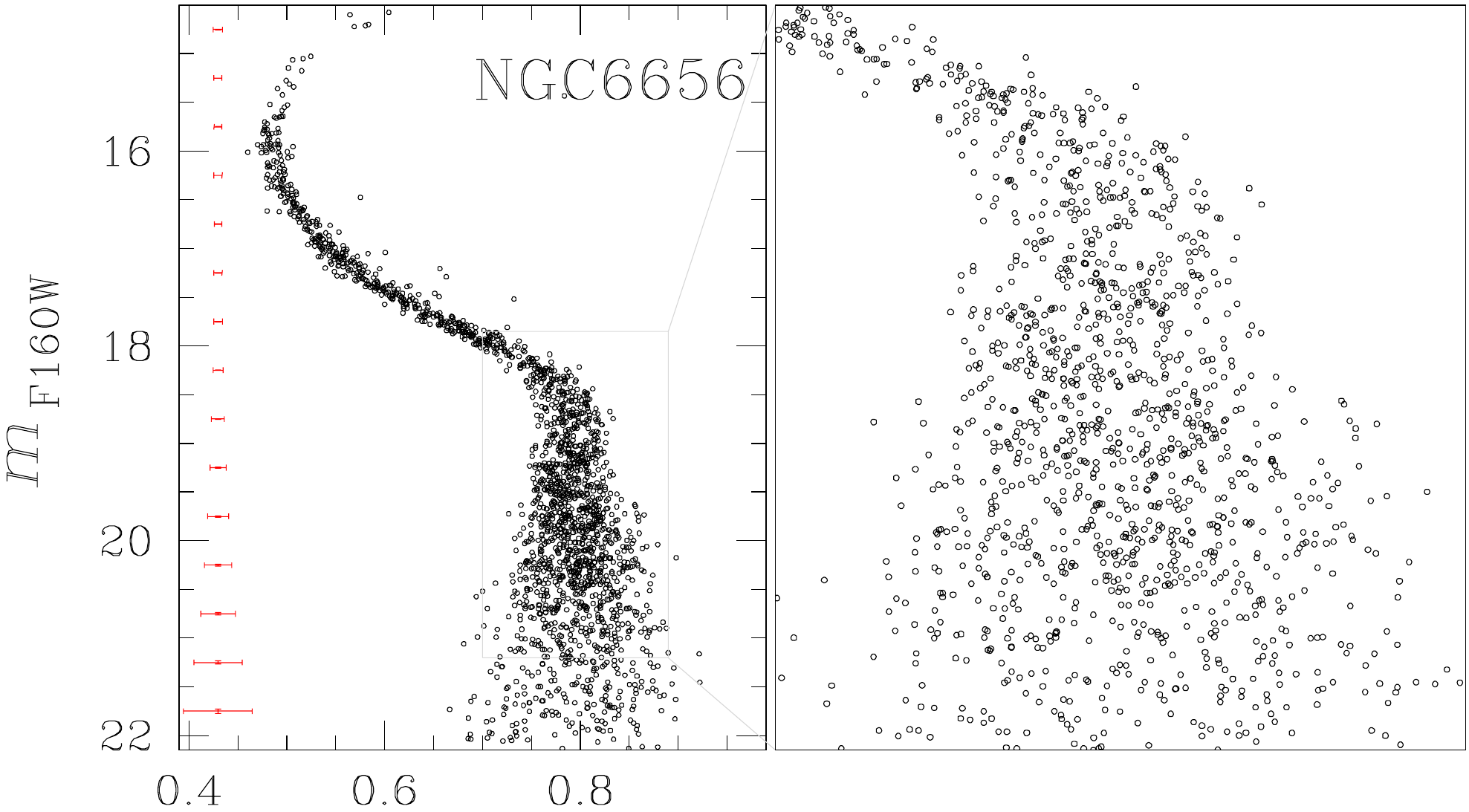}
\includegraphics[width=8.7cm,trim={0.7cm 5.5cm 0.2cm 11.9cm},clip]{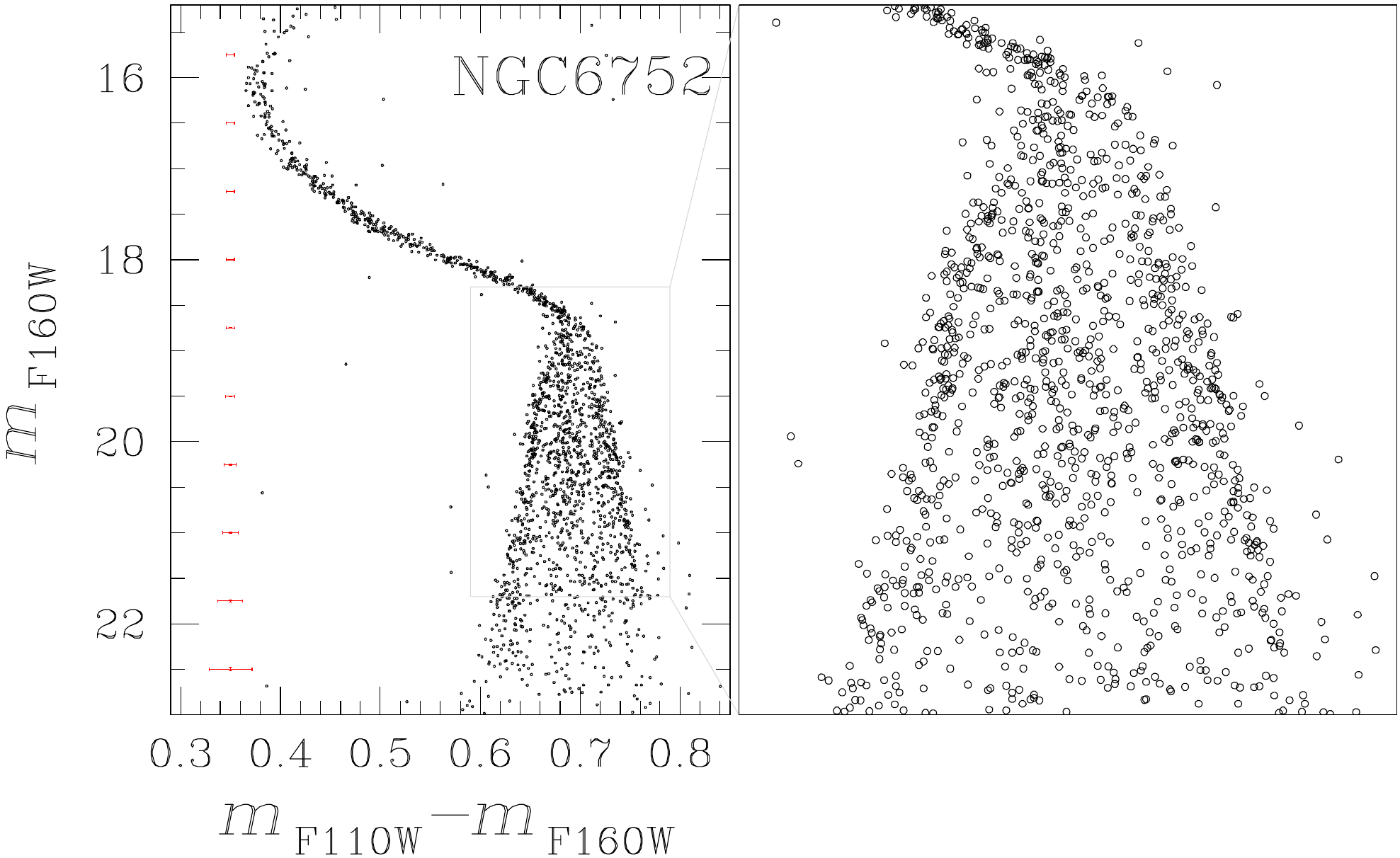}
\includegraphics[width=8.7cm,trim={0.7cm 5.5cm 0.2cm 11.9cm},clip]{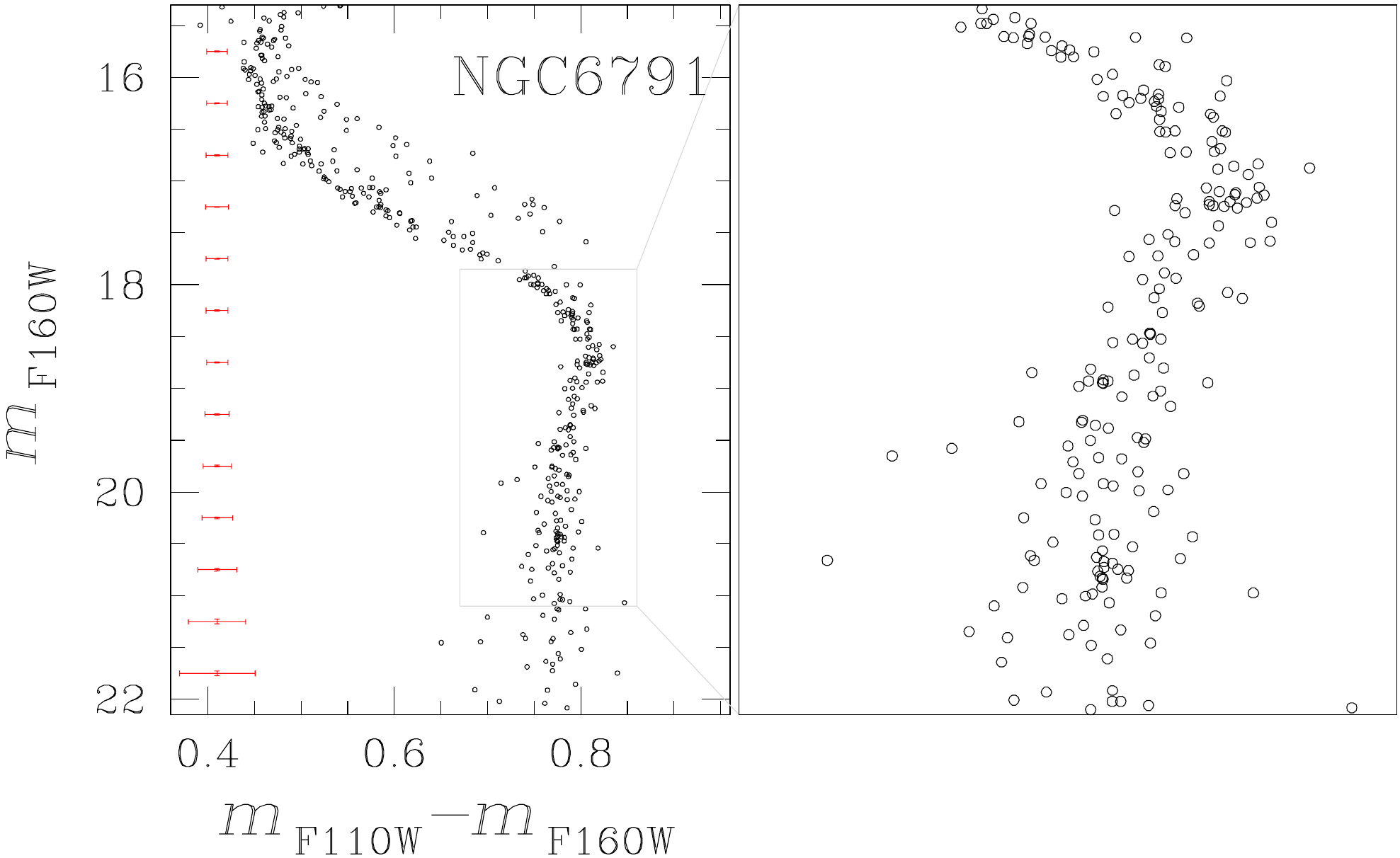}
\caption{Collection of $m_{\rm F160W}$ vs.\,$m_{\rm F110W}-m_{\rm F160W}$ CMDs for the clusters studied in this paper. We show on the right of each CMD a zoom around the MS knee. Red bars represent the color uncertainties at different $m_{\rm F160W}$ levels.}
\label{fig:NIRcmd}
\end{figure*}

The resulting $m_{\rm F160W}$ vs. $m_{\rm F110W} - m_{\rm F160W}$ CMDs are plotted in Figure~\ref{fig:NIRcmd}, where we also show a zoom around the MS region below the knee (i.e. the saddle at $\sim$2$-$3 magnitudes below the MS turn-off). Clearly, below the knee, the MS width of all studied GCs is wider than the color spread expected by photometric errors alone, thus demonstrating that all GCs host MPs.
In contrast, the color broadening of M-dwarfs in NGC\,6791 is consistent with what we expect from observational uncertainties, indicating that the CMD of this open cluster is consistent with a simple population.

\begin{figure*}[ht]
\begin{center} 
  \includegraphics[width=13cm,trim={0.7cm 5cm 0.2cm 6cm},clip]{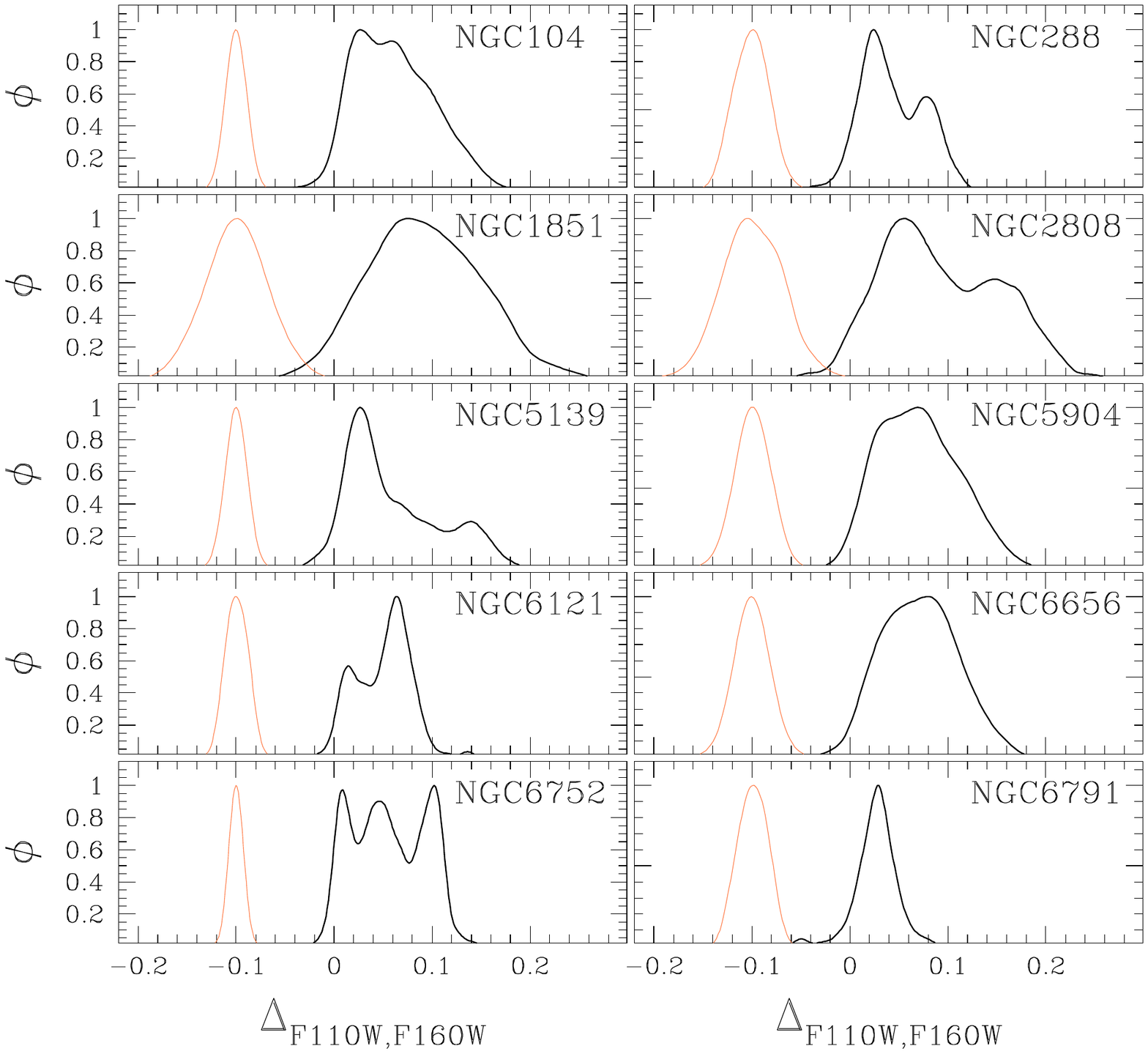}
  \caption{$\Delta_{\rm F110W,F160W}$ kernel-density distributions for M-dwarfs in the F160W magnitude interval between 0.5 and 2.5 mag below the MS knee (black). Orange curves indicate the corresponding distributions of observational errors. For clearness, the error distributions are shifted by $-$0.1 mag in $\Delta_{\rm F110W,F160W}$. }
 \label{fig:KDS} 
 \end{center} 
\end{figure*}

A visual inspection of the CMDs in Figure~\ref{fig:NIRcmd} reveals that the color distribution of M dwarfs significantly changes from one cluster to another.
The GCs NGC\,288, NGC\,2808 and M\,4 exhibit bimodal MSs, and a triple MS is clearly visible in NGC\,6752. On the contrary, the M-dwarfs of the remaining GCs show more-continuous color distributions.
To further highlight the variety of the MPs phenomenon among VLM stars, for each cluster we selected MS stars in the F160W magnitude interval between 0.5 and 2.5 mag below the MS knee (from Lagioia et al.\,in preparation). The corresponding $m_{\rm F160W}$ vs.\,$m_{\rm F110W} - m_{\rm F160W}$ CMD is then verticalized as in \citet[][see their Section 3.2]{milone2017a} to derive the $\Delta_{\rm F110W,F160W}$ pseudo color. The distribution of this quantity, plotted in Figure~\ref{fig:KDS}, corroborates the idea of a variety in MP patterns in our GCs sample.

To quantify the MS color width, we considered M dwarfs in a $\pm$0.2-mag wide interval located 2.0 F160W magnitudes below the MS knee.
We first defined the quantity, $W^{\rm obs}_{\rm F110W,F160W}$, 
which is indicative of the observed MS width and corresponds to the difference between the 96$^{\rm th}$ and the 4$^{\rm th}$ percentile of the F110W-F160W color distribution of the selected M-dwarfs. Then, to estimate the intrinsic MS width, $W_{\rm F110W,F160W}$, we subtracted in quadrature the contribution of observational errors from $W^{\rm obs}_{\rm F110W,F160W}$. The error associated to $W_{\rm F110W,F160W}$ has been determined by bootstrapping with replacements over the sample of M-dwarfs, then repeated 1,000 times. The derived errors refer to one standard deviation of the bootstrapped measurements.

\begin{figure*}[htp]
\begin{center} 
  \includegraphics[width=17cm]{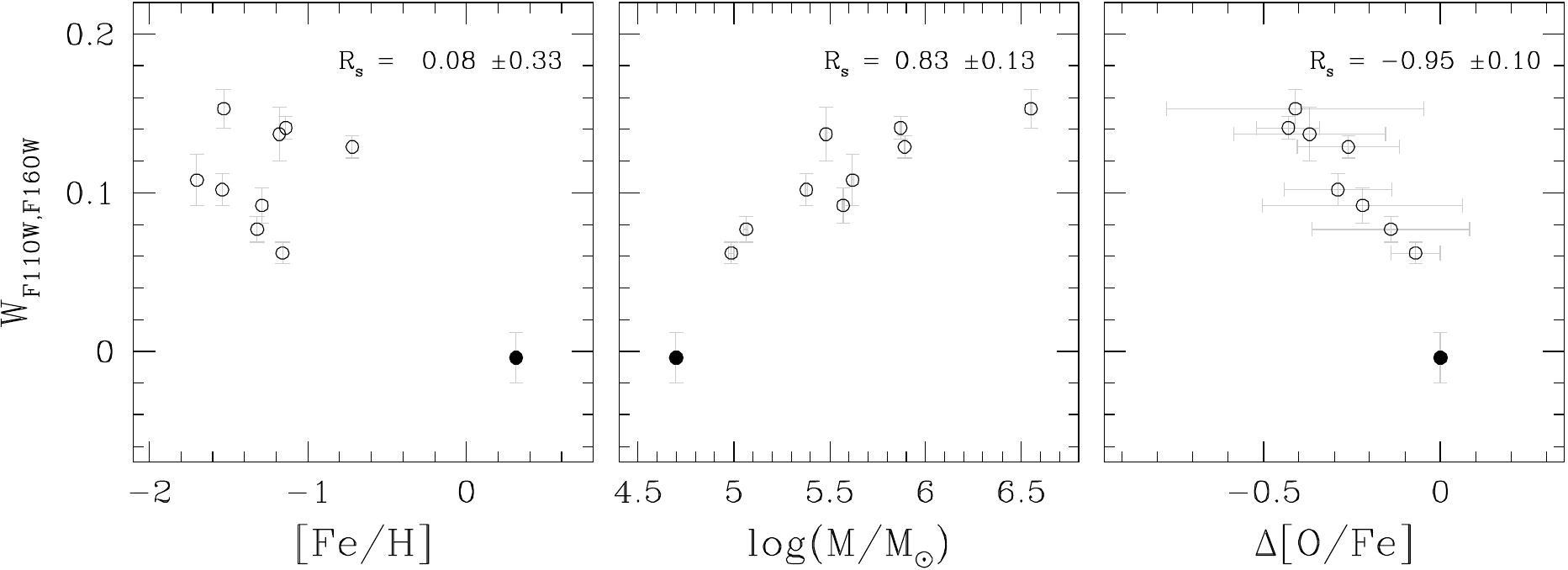}
  \caption{The MS width $W_{\rm F110W,F160W}$, is plotted against cluster metallicity (left), logarithm of cluster mass (middle) and average oxygen difference between of 2G and 1G (right). The black dots indicate the NGC\,6791 measurements. The Spearman's rank correlation coefficients for GC measurements are quoted in each panel.}
 \label{fig:cors} 
 \end{center} 
\end{figure*}

We find that $W_{\rm F110W,F160W}$ ranges from $\sim$0.06 to 0.15 mag in the studied GCs and is consistent with zero in NGC\,6791. As shown in Figure~\ref{fig:cors}, $W_{\rm F110W,F160W}$ does not correlate with cluster metallicity  \citep[][2010 version]{harris1996a, villanova2018} but significantly correlates with cluster mass \citep[from][]{platais2011, baumgardt2018} and anticorrelates with the oxygen difference between GC 2G and 1G stars by ~\citet[][]{marino2019a} and based on high-resolution spectroscopy. 

\section{Multiple populations in \texorpdfstring{NGC\,2808}{} and \texorpdfstring{M\,4}{}} \label{sec:selection}

In the following, we investigate in detail two GCs that have been widely studied in the context of MPs: NGC\,2808 and M\,4.
\begin{itemize}
    \item NGC\,2808 is one of the most-complex clusters in the context of MP patters and is the first GC, after $\omega$\,Cen, where stellar populations with extreme helium abundances have been detected both from photometry and spectroscopy \citep[e.g.][]{dantona2005, piotto2007, marino2014a}. Moreover, it is also the GCs where it has been first discovered a double MS of VLM stars \citep[][]{milone2014}. 
     The pseudo two-color diagram called chromosome map (ChM) of its RGB and upper MS reveals at least five distinct stellar populations (called A, B, C, D, and E), with different helium abundances: populations A, B and C exhibit nearly pristine helium contents, while populations D and E are strongly helium enhanced, up to Y$\sim$0.31 and Y$\sim$0.36, respectively \citep[][]{milone2015}. Helium-rich populations also exhibit extreme content  of various light elements including C, N, O, Na, Mg, Al, Si and K \citep[see][ and references therein, for details on the chemical composition of stellar populations in NGC\,2808]{carretta2009a,  carretta2018a, carretta2015a, mucciarelli2015a, marino2017a, marino2019a, latour2019a}. 
     
     \item In contrast, M\,4 hosts two distinct stellar populations with moderate differences in C, N, O, Na and Al \citep[e.g.][]{marino2008a, marino2011a, marino2017a, carretta2009a, villanova2011a} and similar helium abundances \citep[e.g.][]{milone2018, lagioia2018a, tailo2019}. The distinct groups of 1G and 2G stars have been identified along the entire CMD, from the asymptotic and horizontal branches \citep[e.g.][]{marino2011a, marino2017a} to the VLM regime \citep[][]{milone2014}.
\end{itemize}

In this work, we derive the LFs and the MFs of MPs in both clusters along a wide range of stellar masses, from the brightest part of the MS to the VLM regime\footnote{Due to the small radial sampling of NIR/WFC3, we did not investigate on any radial variation of the MFs inside the FoV. For that, the LFs and MFs of NGC\,2808 and M\,4 derived in the following Sections are referred to all stars in a given FoV.}.
These two clusters have been chosen for three reasons. First, as discussed above, their MPs have been already identified and chemically characterized both between the MS turn-off and knee (hereafter upper MS) and below the MS knee (hereafter lower MS). 
 Hence, it is possible to connect MPs below and above the MS knee. 
Second, our dataset allows us to identify the distinct stellar populations, both below the MS knee and along the upper MS, in the same FoV, so that we can compare the phenomenon in the two stellar mass regimes at the same radial distance from GC centre.
Finally, the F110W$-$F160W color distribution of M dwarfs is bimodal, thus allowing us to separate the majority of stars of each population and derive the corresponding MFs. 

For both clusters, we performed AS tests to evaluate the completeness level at different magnitudes. All the stars used to measure the LFs and the MFs of different stellar populations have a completeness $\gtrsim 60 \%$.

\begin{figure*}[htp]
\begin{center} 
  \includegraphics[width=8.0cm,trim={0cm -0.25cm 0cm 0cm},clip]{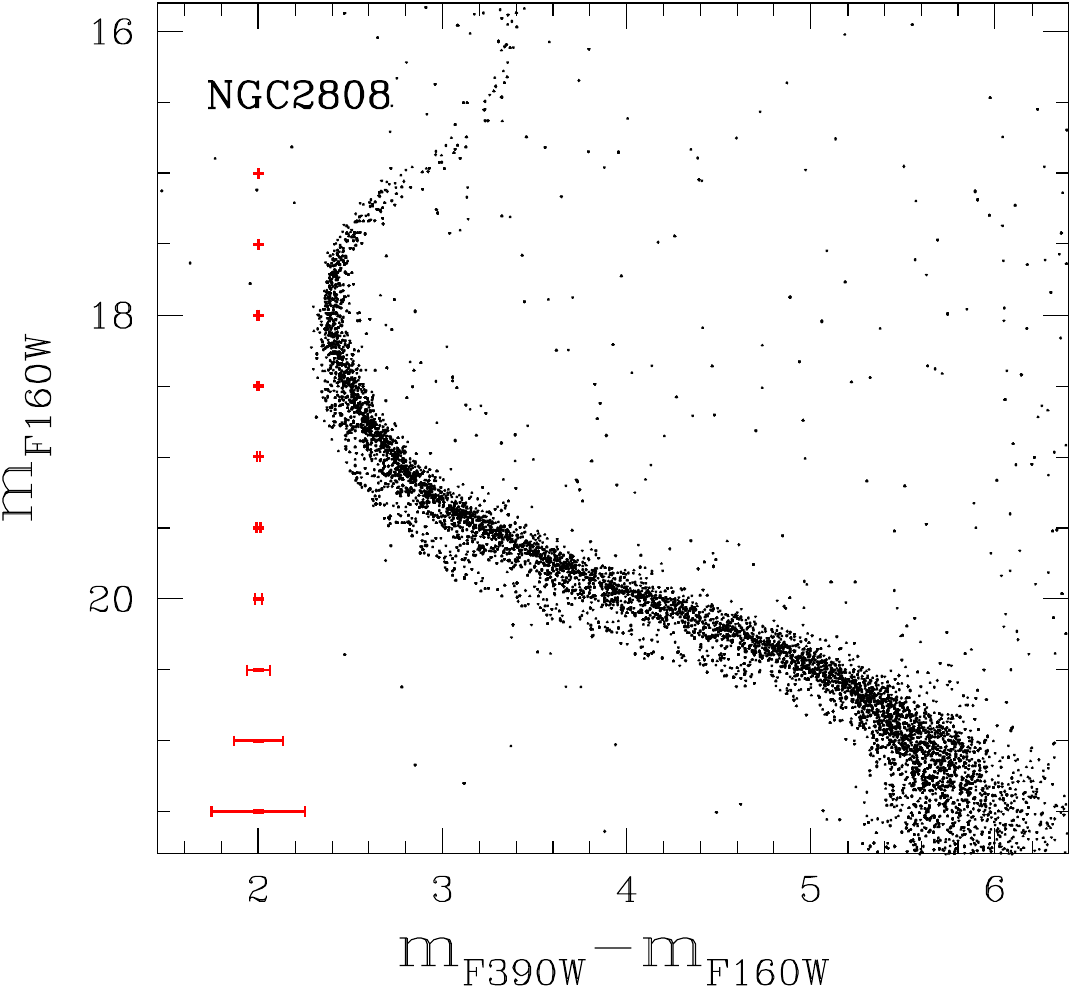}
  \includegraphics[width=8.0cm,trim={0cm 0cm 0cm 0cm},clip]{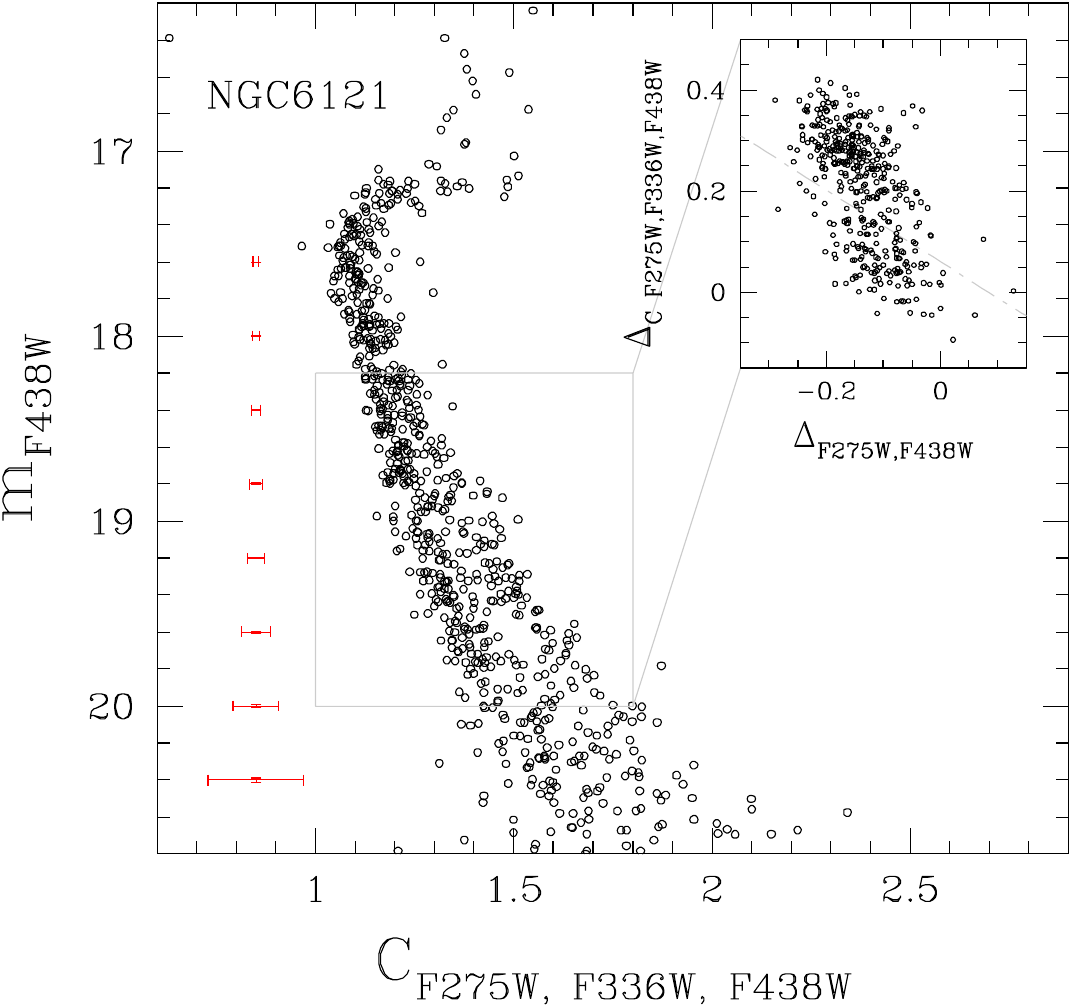}
  \caption{$m_{\rm F160W}$ vs.\,$m_{\rm F390W} - m_{\rm F160W}$ CMD of NGC\,2808 (left panel) and $m_{\rm F438W}$ vs. $C_{\rm F275W,F336W,F438W}$ diagram of M\,4 (right panel). The inset in the right panel shows the ChM for stars inside the grey box.
  Red bars indicate the photometric uncertainties at different magnitude levels.}
 \label{fig:highms} 
 \end{center} 
\end{figure*}

To identify MPs along the upper MS of NGC\,2808, we exploited the $m_{\rm F160W}$ vs.\,$m_{\rm F390W} - m_{\rm F160W}$ CMD (left panel of Figure~\ref{fig:highms}), which allows us to separate three sequences between $19.0\lesssim m_{\rm F160W}\lesssim20.5$. 

For M\,4, we show in the right panel of Figure~\ref{fig:highms} the $m_{\rm F438W}$ vs.\,$C_{\rm F275W,F336W,F438W}$\footnote{$C_{\rm F275W,F336W,F438W} = (m_{\rm F275W}-m_{\rm F336W}) - (m_{\rm F336W}-m_{\rm F438W})$} diagram, where two sequences can be disentangled at $18.2\lesssim m_{\rm F438W}\lesssim20.0$. We used this diagram, along with the $m_{\rm F438W}$ vs.\,$m_{F275W}-m_{F438W}$ CMD, to build the ChM \citep[][]{milone2015, milone2017a}.
In a nutshell, we computed the 4$^{\rm th}$ and the 96$^{\rm th}$ percentiles of the $m_{\rm F275W}-m_{\rm F438W}$ color and $C_{\rm F275W,F336W,F438W}$ pseudo-color distributions of MS stars in different 0.2-wide magnitude bins.  These values have been associated with the median magnitude of stars in each bin and have been linearly interpolated to derive the red and blue boundaries of  the MS. 
Finally, we exploited the boundaries of $m_{\rm F275W}-m_{\rm F438W}$ and $C_{\rm F275W,F336W,F438W}$ to derive the ChM coordinates ($\Delta_{\rm F275W,F438W}$ and $\Delta_{\rm C F275W,F336W,F438W}$) by using the transformations by \citet[][see their Section 3.2]{milone2017a}. The resulting ChM is shown in the inbox in the right panel of Figure~\ref{fig:highms}, and allows us to disentangle the bulk of 1G stars (which are clustered near the origin of the reference frame) from the 2G stars (which exhibit higher values of $\Delta_{\rm C F275W,F336W,F438W}$).
In the next Sections, we will combine information from the diagrams of Figure \ref{fig:highms} and from the $m_{\rm F160W}$ vs.\,$m_{\rm F110W}-m_{\rm F160W}$ CMDs of Figure \ref{fig:NIRcmd} to disentangle MPs and infer their MFs.

\section{Luminosity and Mass Functions of multiple populations in \texorpdfstring{NGC\,2808}{}} \label{sec:lf}

To derive the LFs and the MFs of the distinct populations of NGC\,2808 we analyzed MS stars in different F160W magnitude intervals, separately. Specifically, we identified a sample of upper MS stars with $19.0<m_{\rm F160W}<20.2$ and a group of lower MS stars with $21.0<m_{\rm F160W}<22.5$. This choice is due to the fact that different photometric diagrams are needed to disentangle MPs along the upper  and lower MS.
LFs are obtained by extending to NGC\,2808 the methods by \citet{milone2012b} as discussed in Sections \ref{sub:uMSn2808} and \ref{subsec:lowMSn2808} for the upper- and lower-MS samples, respectively. LFs are converted into MFs in Section \ref{subsec:MFn2808} by using appropriate mass-luminosity relations.

\subsection{Luminosity functions of multiple populations along the upper MS}\label{sub:uMSn2808}

\begin{figure*}[ht]
\begin{center} 
  \includegraphics[height=15cm,clip]{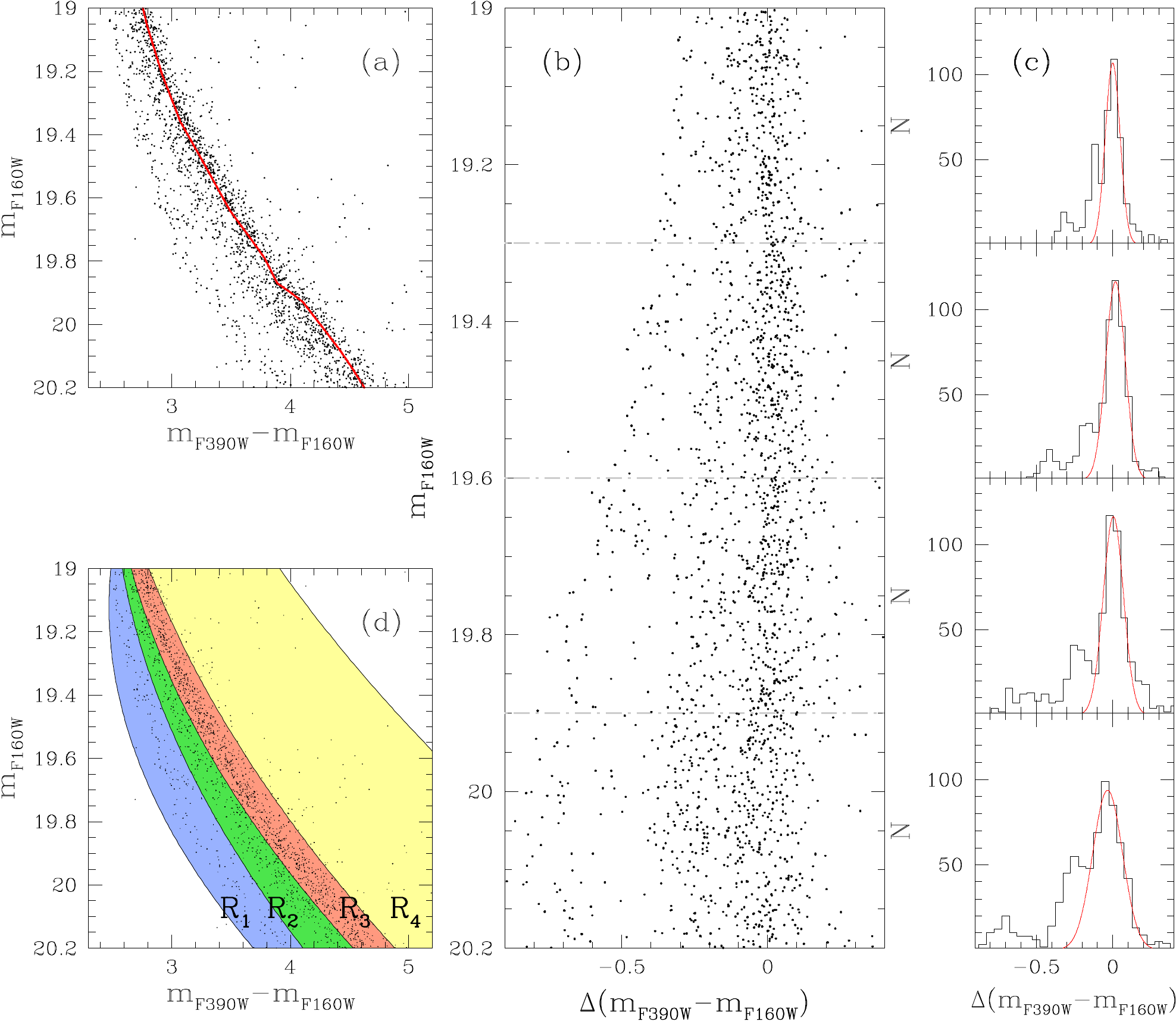}
  \caption{\textit{Panel (a):} $m_{\rm F160W}$ vs.\,$m_{\rm F390W} - m_{\rm F160W}$ CMD of all stars in the three fields of NGC\,2808. The red line shows the first-guess fiducial line of the rMS. \textit{Panel (b):} $m_{\rm F160W}$ vs.\,$\Delta(m_{\rm F390W} - m_{\rm F160W})$ verticalized diagram obtained from the rMS fiducial line (see text for details). \textit{Panel (c):} histogram distribution of $\Delta(m_{\rm F390W} - m_{\rm F160W})$ in four different magnitude bins. The red lines represent the best-fit Gaussian functions of the rMS stars. \textit{Panel (d):} regions R$_{\rm 1}$, R$_{\rm 2}$, R$_{\rm 3}$ and R$_{\rm 4}$ in the $m_{\rm F160W}$ vs.\,$m_{\rm F390W} - m_{\rm F160W}$ CMD, colored in blue, green, red, and yellow, respectively.}
 \label{fig:met} 
 \end{center} 
\end{figure*}

To investigate the upper MS of NGC\,2808 we exploited the $m_{\rm F160W}$ vs.\,$m_{\rm F390W} - m_{\rm F160W}$ CMD shown in panel (a) of Figure~\ref{fig:met}. 
This diagram clearly reveals the triple MS, which is a distinctive feature of NGC\,2808 and has been associated with three stellar populations with different helium abundances. The red MS (rMS) is composed of stars with nearly pristine helium content, whereas the middle and the blue MSs (mMS and bMS) are highly enhanced in helium \citep[up to Y$\sim$0.31 and Y$\sim$0.36, respectively, e.g.][]{dantona2005, piotto2007, milone2012b, milone2015}.
In the following, we derive the LFs and the MFs of the three main populations of NGC\,2808 by extending to the F390W and F160W photometry the procedure introduced by \citet[][]{milone2012b} and illustrated in Figure~\ref{fig:met}.  
We limited our analysis to the $19.0 < m_{\rm F160W} < 20.2$ interval, where the three sequences are more clearly distinguishable.

\begin{itemize}
    \item The first step consists in deriving the fiducial line of each MS and is illustrated in Figure~\ref{fig:met} for the rMS. We initially derived by hand a first-guess fiducial line and calculated the color residuals, $\Delta(m_{\rm F390W} - m_{\rm F160W})$, defined as the difference between the color of each star and the color of the fiducial line at the same F160W magnitude. 
    The verticalized $m_{\rm F160W}$ vs.\,$\Delta(m_{\rm F390W} - m_{\rm F160W})$ diagram is plotted in panel (b) and is used to derive the histogram distributions of $\Delta(m_{\rm F390W} - m_{\rm F160W})$ of MS stars in four 0.3 magnitude bins (panel (c)). 
    Each histogram reveals three main peaks that correspond to the three MSs. We fitted the histogram distribution of rMS stars with a Gaussian function by means of least squares. The best-fit Gaussian is represented in red. Finally, we associated the Gaussian centers with the average magnitude of star in each magnitude bin and linearly interpolated these points to derive the rMS fiducial line. The same procedure has been extended to the mMS and the bMS. 
    
    \item We defined four regions in the CMD, namely R$_{\rm 1}$, R$_{\rm 2}$, R$_{\rm 3}$ and R$_{\rm 4}$ by arbitrarily shifting the three MS fiducial lines. These regions are marked with different colors as illustrated in panel (d) of Figure~\ref{fig:met}. Region R$_{\rm 1}$ is mostly populated by bMS stars and its blue and red boundaries are obtained by subtracting and adding to the bMS fiducial line 3$\times \sigma_{\rm bMS}$, respectively.
    Similarly, the red and blue boundaries of the region associated with the rMS, R$_3$, are derived by subtracting 1$\times \sigma_{\rm rMS}$ and adding 3$\times \sigma_{\rm rMS}$ to the rMS fiducial line, respectively.
    Here, $\sigma_{\rm bMS}$ and $\sigma_{\rm rMS}$ are the color dispersions of the bMS and the rMS, respectively, and are derived from the best-fit Gaussian functions.
    The region R$_{\rm 2}$ is placed between R$_{\rm 1}$ and R$_{\rm 3}$ and is mainly populated by mMS stars.
    Finally, the region R$_{\rm 4}$ is adjacent to R$_{\rm 3}$ and is mostly occupied by binary stars composed of two MS stars. Its red boundary corresponds to line of equal-mass rMS-rMS binaries red-shifted by 3$\times \sigma_{\rm rMS}$.

\item
The last step consists in deriving the number of stars of each population in different magnitude bins.
It is worth to notice that to maximize the sample of stars and obtaining robust results, the fiducial lines and the regions of the CMD have been derived from all stars in NGC\,2808, combining Field A, B and C. On the contrary, the LFs have been estimated for stars in each field, separately.

Clearly,  as a consequence of photometric uncertainties, each region has a net contamination from stars belonging to different populations. To account for this contamination effects, and to derive the real number of stars of each population we exploited the following procedure.
Specifically,  the number of stars, $N_{\rm i}$, which we observed falling in a region $R_{\rm i}$ is: 

\begin{equation}\label{eq}
    \resizebox{.88\hsize}{!}{$N_{\rm i} = N_{\rm bMS}f_{\rm i}^{\rm bMS} + N_{\rm mMS}f_{\rm i}^{\rm mMS} + N_{\rm rMS}f_{\rm i}^{\rm rMS} + f^{\rm BIN}N_{\rm MS}f_{\rm i}^{\rm BIN}$ }
\end{equation}
where $N_{\rm bMS}$, $N_{\rm mMS}$ and $N_{\rm rMS}$ are the numbers of stars of the three populations, $N_{\rm MS}$ their sum, $f_{\rm i}^{\rm bMS}$, $f_{\rm i}^{\rm mMS}$ and $f_{\rm i}^{\rm rMS}$ the fraction of stars of the three populations that fall in the i$^{\rm th}$ region. The last term of the equation accounts for the presence of binary systems. In particular, $f_{\rm i}^{\rm BIN}$ indicates the fraction of binary stars that populated the i$^{\rm th}$ region of the CMD and $f^{\rm BIN}$ is the total binary fraction. Having four regions, we need to solve four equations to derive the number of stars in each population and the total binary fraction.
 
The values of $f_{\rm i}^{\rm bMS}$, $f_{\rm i}^{\rm mMS}$ and $f_{\rm i}^{\rm rMS}$ are inferred from simulated CMDs made of ASs (see Section~\ref{subsec:as}). We simulated 40,000 ASs for each population, disposed along the corresponding fiducial line. The values of $f_{\rm i}^{\rm BIN}$ are calculated as the fraction of binary stars in the corresponding region, with respect to the total number of binaries.

The degeneracy between the LFs and the fraction of binaries provides the main challenge to estimate $f_{\rm i}^{\rm BIN}$. To break this degeneracy, we adopted the following iterative procedure. At the first iteration, we fixed $f^{\rm BIN} = 0$ and solved the system of equations~\ref{eq}, thus finding first estimates of $N_{\rm bMS}$, $N_{\rm mMS}$ and $N_{\rm rMS}$. 
  
Then, we simulated a CMD composed of MS-MS binaries alone, by assuming that each population hosts the same binary fraction. 
We adopted a flat mass-ratio distribution for binaries and enhanced by ten times the numbers of bMS-bMS, mMS-mMS, and rMS-rMS binaries, to increase the statistics.
The resulting CMD has been used to improve the estimates of $f_{\rm i}^{\rm BIN}$. The last step of the first iteration consists in solving equations~\ref{eq} and deriving $N_{\rm bMS}$, $N_{\rm mMS}$, $N_{\rm rMS}$ and $f^{\rm BIN}$.

In the subsequent iterations, we used the best estimates of $N_{\rm bMS}$, $N_{\rm mMS}$ and $N_{\rm rMS}$ as input for generating a CMD composed of binaries alone to improve the values of $f_{\rm i}^{\rm BIN}$, and then we solved the system of equations~\ref{eq}. 
We repeated this step until the $f^{\rm BIN}$ value changed by less than 0.001 between two subsequent iterations. 
The total binary fraction obtained is $0.033 \pm 0.020$, $0.064 \pm 0.035$ and $0.082 \pm 0.036$ in fields A, B and C, respectively.

The number of stars in the various magnitude intervals, corrected for completeness, provide the LFs of the three populations of NGC\,2808. Noticeably, the star counts are provided in units of magnitude and area to remove the dependence from the bin size and the area of the analyzed FoV. The LFs of the three populations in the upper MS of NGC\,2808 are represented, for the three fields combined, in Figure~\ref{fig:3ms}.
\end{itemize}

\begin{figure} 
\begin{center} 
  \includegraphics[height=8.5cm,clip]{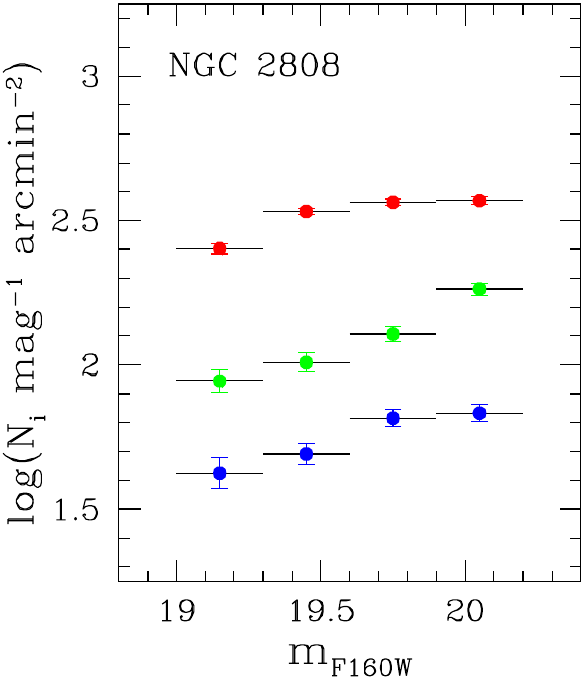}
  \caption{Luminosity functions of the three MSs of all stars in NGC\,2808. Red, green and blue dots represent rMS, mMS and bMS stars, respectively. \\}
 \label{fig:3ms} 
 \end{center} 
\end{figure} 

To estimate the uncertainty on the star counts used to derive the LF, we simulated 1,000 CMDs that are composed of three stellar populations distributed along the fiducials of NGC\,2808 affected by the same photometric errors as real stars. The total number of simulated stars corresponds to the number of observed stars and the fraction of stars in the simulated blue- middle- and red MS matches the observed ones.
We applied to each simulated CMD the procedure described above and estimate the corresponding LFs. The uncertainty associated to each LF bin is provided by the standard deviation of the corresponding 1,000 LF determinations.

\subsection{Luminosity functions of multiple populations along the lower MS} \label{subsec:lowMSn2808}

\begin{figure*} 
\begin{center} 
  \includegraphics[height=7.5cm,clip]{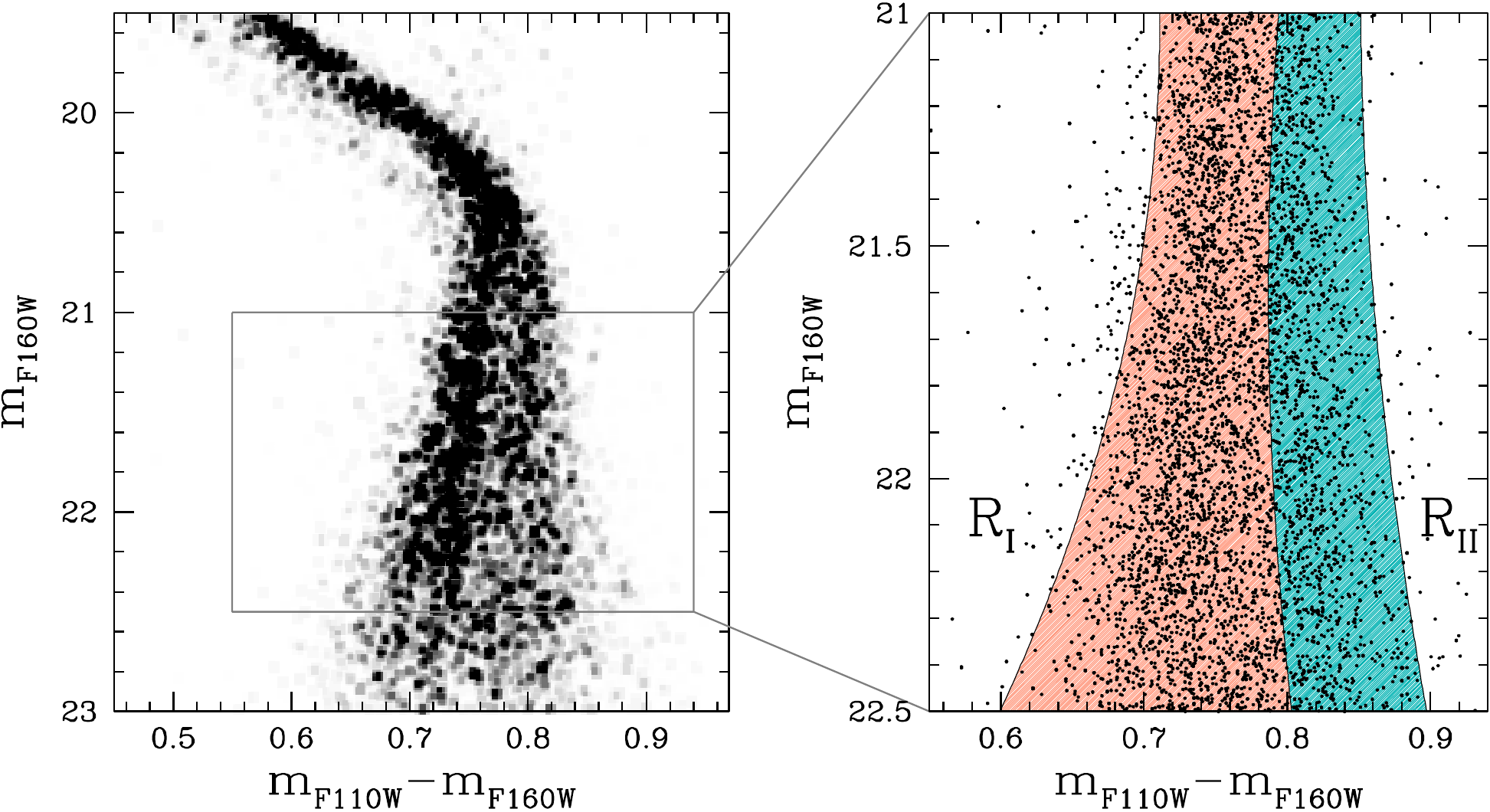}
  \caption{{\it{Left panel:}} Hess diagram of the $m_{\rm F160W}$ vs.\,$m_{\rm F110W} - m_{\rm F160W}$ CMD for all the stars in NGC\,2808. {\it{Right panel:}} CMD of stars inside the black box. Black lines show the boundaries of regions $R_{\rm I}$ and $R_{\rm II}$, colored in red and azure, respectively. \\}
 \label{fig:ir} 
 \end{center} 
\end{figure*}

To measure the LF of MPs along the lower MS of NGC\,2808 we exploited the $m_{\rm F160W}$ vs. $m_{\rm F110W} - m_{\rm F160W}$ CMD (see Figure~\ref{fig:ir} for the corresponding Hess diagram).
As demonstrated by \citet[][]{milone2012a}, the MS with blue $m_{\rm F110W} - m_{\rm F160W}$ color, hereafter MS-I, is the counterpart of the rMS, whereas the  mMS and the bMS merge together into the red MS (MS-II). 
In fact, the color of upper MS stars in NGC\,2808 is dominated by the effect of helium, which makes a star hotter (bluer) when its abundance increases. Moving to the lower MS, two main mechanisms operate: the increase in radiative opacity and the Collision Induced Absorption (CIA) of the H$_{\rm 2}$ molecule, which make the stars redder and bluer, respectively. When moving through lower masses, the CIA dominates in MS-I, which becomes bluer. 
MS-II has a bigger helium amount that shift its color through blue, but its lower H abundance makes the CIA contribution drop, so that the increase of opacity dominates and its color becomes redder. In the NGC\,2808 VLM stars, these two effects almost compensate each other when considering the F110W$-$F160W color, making it roughly unaffected by helium variations. It is instead sensitive to oxygen variations thanks to the F160W band. MS-I stars are enriched in oxygen with respect to MS-II stars, so that they have fainter $m_{\rm F160W}$ and therefore a bluer F110W$-$F160W color.
Unfortunately, the present dataset does not allow us to disentangle between mMS and bMS below the MS knee.

To derive the LF of MS-I and MS-II stars, we adapted the procedure of Section \ref{sub:uMSn2808} to the NIR CMD region between $m_{\rm F160W}$=21.0 and 22.5 highlighted by the box in the Hess diagram, where the split MS is clearly visible.
As illustrated in the right panel of Figure~\ref{fig:ir}, we defined the regions $R_{\rm I}$ and $R_{\rm II}$ in the CMD, centered on MS-I and MS-II stars and colored in red and azure, respectively. $R_{\rm I}$ is delimited by fiducials of MS-I stars shifted by 2$\times \sigma_{\rm MS-I}$ to the blue and 1$\times \sigma_{\rm MS-I}$ to the red (black lines), while the red boundary of $R_{\rm II}$ corresponds to the MS-II fiducial red-shifted by 2$\times \sigma_{\rm MS-II}$.
Here, $\sigma_{\rm MS-I}$ and $\sigma_{\rm MS-II}$ correspond to the color broadening of MS-I and MS-II stars, respectively, and are derived as in Section \ref{sub:uMSn2808}. 
 
Since the MSs run almost vertical in the analyzed region of the $m_{\rm F160W}$ vs.\,$m_{\rm F110W} - m_{\rm F160W}$ CMD, binary stars are nearly mixed with single stars. Hence, we did not account for binaries and derived the numbers of MS-I and MS-II stars in 0.3 F160W magnitude intervals by solving the equation:
 \begin{equation}\label{eq2}
    \resizebox{.45\hsize}{!}{$N_{\rm i} = N_{\rm MS-I}f_{\rm i}^{\rm MS-I} + N_{\rm MS-II}f_{\rm i}^{\rm MS-II}$},
\end{equation}
where $N_{\rm i}$ is the number of stars in the region $R_{\rm i}$ and $N_{\rm MS-I}$ and $N_{\rm MS-II}$ are the numbers of MS-I and MS-II stars, respectively. $f_{\rm i}^{\rm MS-I}$ and $f_{\rm i}^{\rm MS-II}$ are the fractions of stars of each population that fall into $R_{\rm i}$ and are estimated by means of AS tests.

\begin{figure*} 
\begin{center} 
  \includegraphics[height=15.5cm,clip]{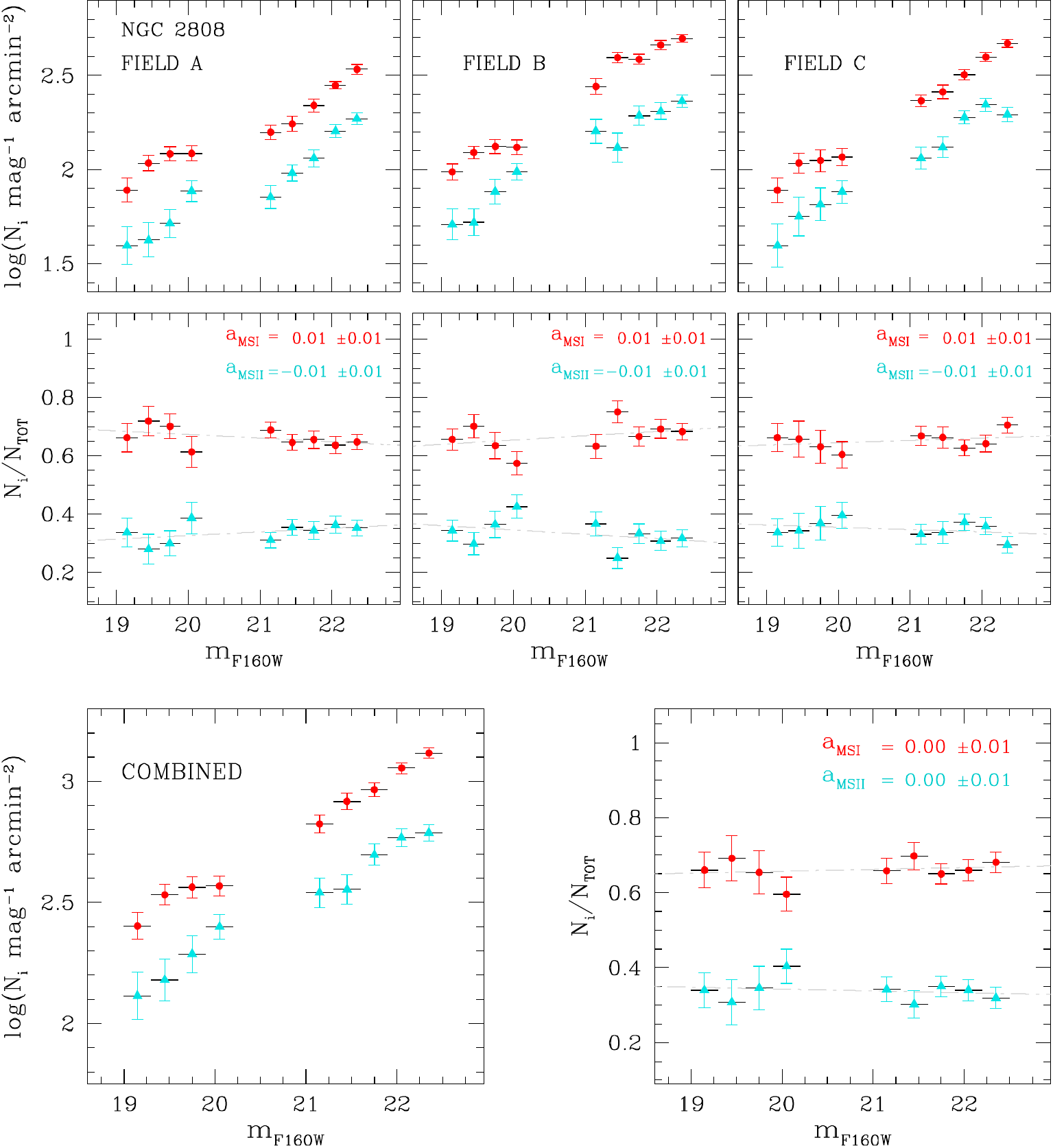}
  \caption{\textit{Upper and middle panels:} LFs and populations ratios of NGC\,2808 MS-I and MS-II stars (red and cyan dots) in Field A, B, and C. \textit{Lower panels:} LFs and population ratios of multiple populations in NGC\,2808 from all the field. The black horizontal bars associated with each point represent the amplitude of the corresponding magnitude bin. The grey dot-dashed lines are the best-fit straight lines and their slopes are reported in the diagrams.}
 \label{fig:lf2808} 
  \end{center} 
\end{figure*} 

The LF of MS-I and MS-II in Fields A, B and C are displayed in the upper panels of Figure~\ref{fig:lf2808}. They have very similar shape, with the number of stars by unit of area and magnitude bin increasing through fainter $m_{\rm F160W}$. At $m_{\rm F160W} \sim 20.2$, the MS-I LF seems to flatten, but it starts again to increase in the low MS regime.
The middle panels show the population ratios in each field in function of the magnitude bin, obtained by dividing the number of red and azure stars from the LFs to the total number of stars ($N_{\rm TOT}$). 
We then combined the results from the three fields finding the LFs and the population ratios of MS-I and MS-II for all the stars in NGC\,2808 (bottom-left and right panel of Figure~\ref{fig:lf2808}, respectively).
The population ratios have been least-squared fitted with the grey dot-dashed straight lines. Their slopes, reported in each panel, are consistent with a null value, so with a flat dependence with the magnitude.
The black bars highlight the magnitude bin range of each point of the LFs and the population ratios.

Furthermore, we used a p-value test to evaluate the statistical significance of the ratios distribution flatness. Briefly, we simulated 10,000 flat-ratio distributions, testing the null hypothesis that statistical fluctuations produce the observed profile. Each simulation was generated starting from the weighted-average of the empirical ratio distributions and then scattering each point following the observational errors. We measured the deviation from flatness through the chi-square of each simulation ($\chi^{2}_{\rm sim}$) and of the observed profile ($\chi^{2}_{\rm obs}$). Then, we calculated the probability that the null hypothesis is false (the {\it{p}} value) by computing the fraction of simulations such that $\chi^{2}_{\rm sim}>\chi^{2}_{\rm obs}$. The null hypothesis is considered not true if ${\it{p}}<0.05$. We found a {\it{p}} value of 0.80, 0.15, 0.71 and 0.91 for Field A, B, C and their combination, respectively, confirming the flatness of the ratio trends.

By performing a weighted average of the populations ratios in all bins from both the upper and lower MS, we found that the fraction of MS-I and MS-II are 0.67$\pm$0.04 and 0.33$\pm$0.04.

\subsection{Mass functions of multiple populations}\label{subsec:MFn2808}

To measure the MFs of the distinct stellar populations, we converted magnitudes into masses through the mss-luminosity relations by \citet[][]{dotter2008}. 
For NGC\,2808, we used the isochrones from the Dartmouth Stellar Evolution Database that provide the best fit with the observed CMD and correspond to age of 11.5 Gyr \citep[][]{milone2014}, [Fe/H] $= -1.14$ \citep[][2010 version]{harris1996a} and [$\rm{\alpha}$/Fe]$= 0.4$ \citep[][]{dotter2010a}. 
 
We accounted for the fact that the three MSs of NGC\,2808 host stars with different content of helium, which implies that they follow different mass-luminosity relations. Specifically, we adopted Y=0.272, 0.336 and 0.386 for the red, middle and blue MS, respectively. These helium abundances are based on the results from \citet[][]{milone2015}, who inferred the relative helium content of the five RGB MPs. They also found that the blue and middle MSs identified in our paper correspond to the populations E and D, respectively, whereas the red MS is composed of their populations A, B and C that our data cannot distinguish below the knee. 
  
\begin{figure} 
\begin{center} 
 \includegraphics[height=7cm,clip]{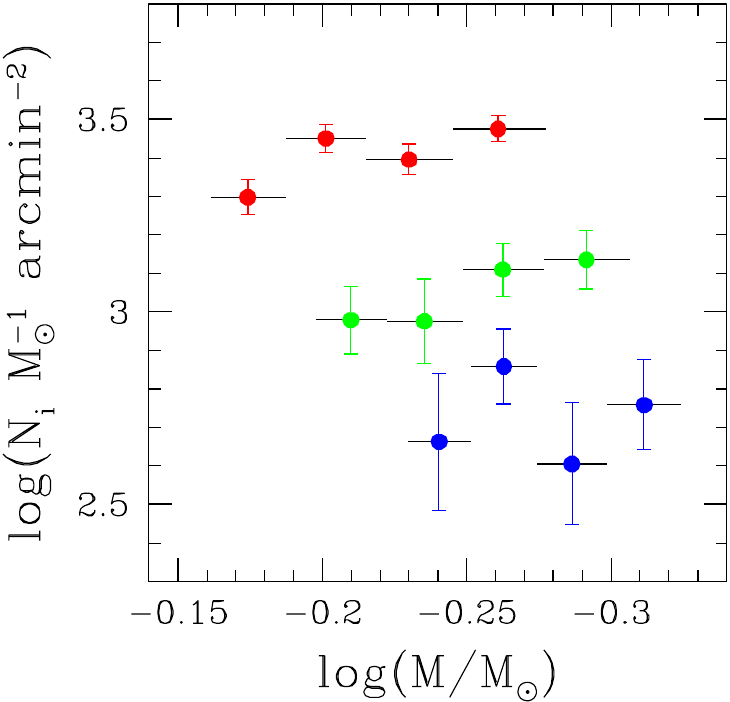}
 \caption{Mass functions of red MS (red), middle MS (green) and blue MS (blue) of NGC\,2808. }
 \label{fig:tr} 
 \end{center} 
\end{figure}

The resulting MFs are illustrated in Figure~\ref{fig:tr}, where we plot the logarithm of the number of rMS, mMS and bMS stars in each magnitude bin normalized per unit mass and unit area against the logarithm of stellar mass.
The small amount of stars in each population (particularly in the bMS) and, more importantly, the very narrow range of masses ($\sim$0.1$\mathcal{M}_{\rm \odot}$) covered by these data do not allow us a meaningful estimate of the slopes of the MF.

\begin{figure} 
\begin{center} 
  \includegraphics[height=12cm,clip]{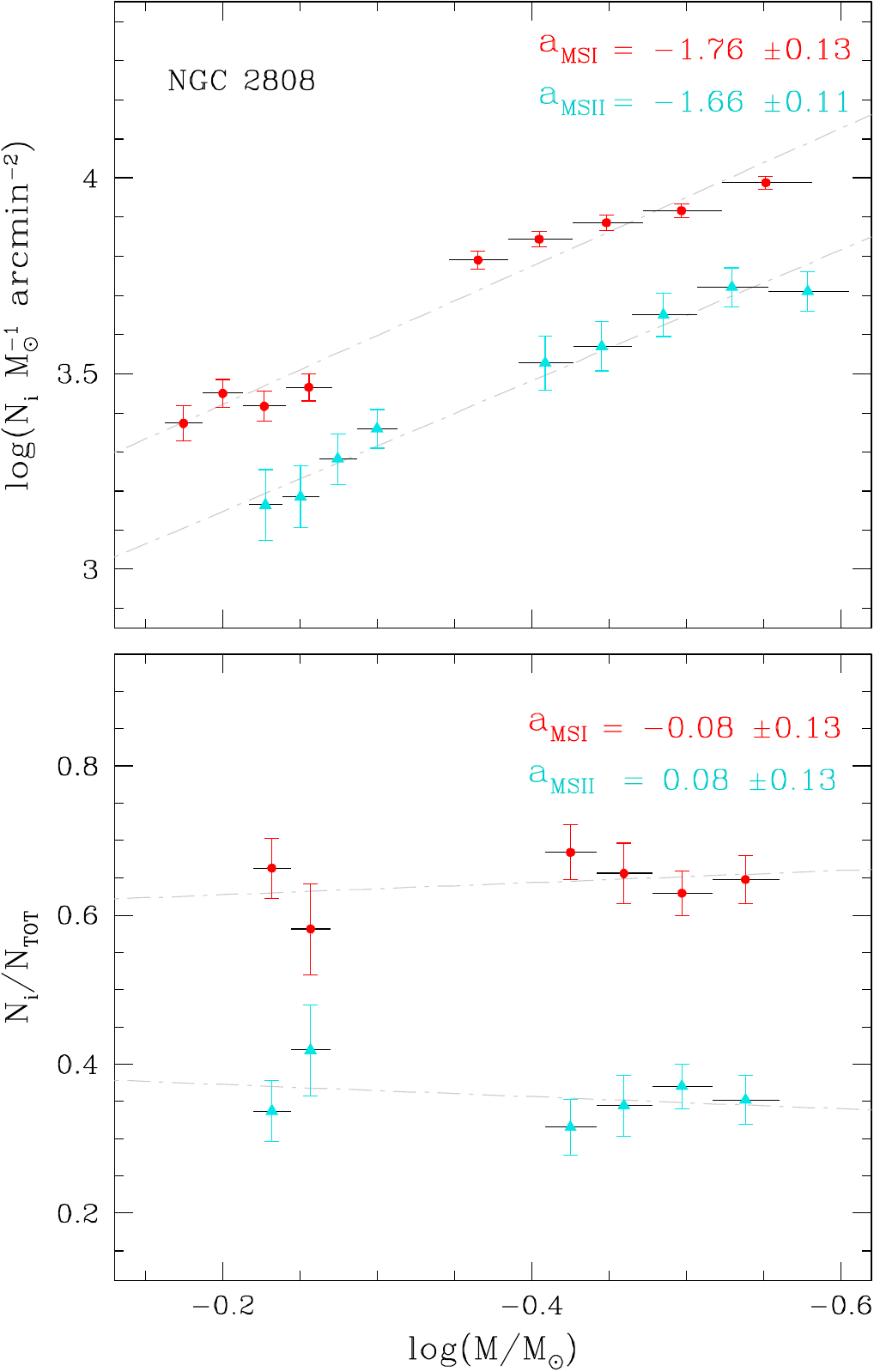}
  \caption{MFs and populations ratios of MS-I and MS-II populations (red and cyan dots) for all NGC\,2808 stars. Best-fit lines are represented with grey dot-dashed lines and their slopes are reported in the diagram. Black bars illustrate the mass extension of each bin.}
 \label{fig:mf2808} 
 \end{center} 
\end{figure}

To investigate the MFs of MPs over a wider interval of stellar masses, we analyzed  the two groups of MS-I stars, which correspond to the red MS and have average helium content Y=0.272, and MS-II stars, which comprise both middle and blue-MS stars and have Y=0.355. 
The resulting MFs are plotted in the  top panels of Figure~\ref{fig:mf2808}. 
We least-squared fitted the MFs by means of straight lines (grey dot-dashed lines) and found that the MF of MS-I and MS-II stars are consistent with having the same slope within one $\sigma$. 

To further compare the MFs of MS-I and MS-II stars, we calculated the fractions of MS-I and MS-II stars in different intervals of stellar mass. As illustrated in the bottom panels of Figure~\ref{fig:mf2808}, the population ratios have a flat trend (with slopes consistent with zero and a {\it{p}}-value equal to 0.84) and average values of 0.65$\pm$0.03 and 0.35$\pm$0.03, consistent with the fractions inferred from the LFs.

MFs are derived by considering mass bins with equal width. As pointed out by \citet[][]{apellaniz2005}, this approach could introduce a bias in their determinations, especially when the number of stars per bin is highly variable. 
To investigate whether our results are affected by this systematic, we repeated our measurements by using equal-number bins. The resulting slopes of MS-I and MS-II resulted in $-$1.81$\pm$0.05 and $-$1.70$\pm$0.15, respectively. These values are very similar to the slopes inferred by using equal-width bins, thus implying that different binning assumptions has a negligible effect on the derived slopes.

\section{Luminosity  and Mass Functions of multiple populations in \texorpdfstring{M\,4}{}}\label{sec:lfM4}

M\,4 hosts two stellar populations, one with primordial chemical composition (1G) and the other with 2G-like abundances, which have been found in both upper and lower MS \citep[e.g.,][]{milone2014, milone2020a}.
We identified them in our FoV by exploiting two distinct diagrams. The upper MS has been analyzed through the ChM, while the $m_{\rm F160W}$ vs. $m_{\rm F110W} - m_{\rm F160W}$ CMD has been exploited to investigate MPs below the MS knee. In the following subsections, we describe the procedure to derive the LF of 1G and 2G stars along the upper and the lower MS in Section~\ref{subsec:uMSM4} and ~\ref{subsec:lMSM4}, and present the corresponding MFs.

\subsection{Upper Main Sequence} \label{subsec:uMSM4}

The ChM was built using the $m_{\rm F438W}$ vs.\,$m_{\rm F275W}-m_{\rm F438W}$ CMD and the $m_{\rm F438W}$ vs.\,$C_{\rm F275W,F336W,F438W}$ pseudo CMD, as explained in Section~\ref{sec:panel}. We considered the stars in the $14.4 < m_{\rm F160W} < 16.2$ interval, where the sequences are more-clearly distinguishable and excluded binaries with high mass ratio, q$\gtrsim$0.2, from the analysis.

\begin{figure}[htp]
\begin{center} 
  \includegraphics[height=7cm,clip]{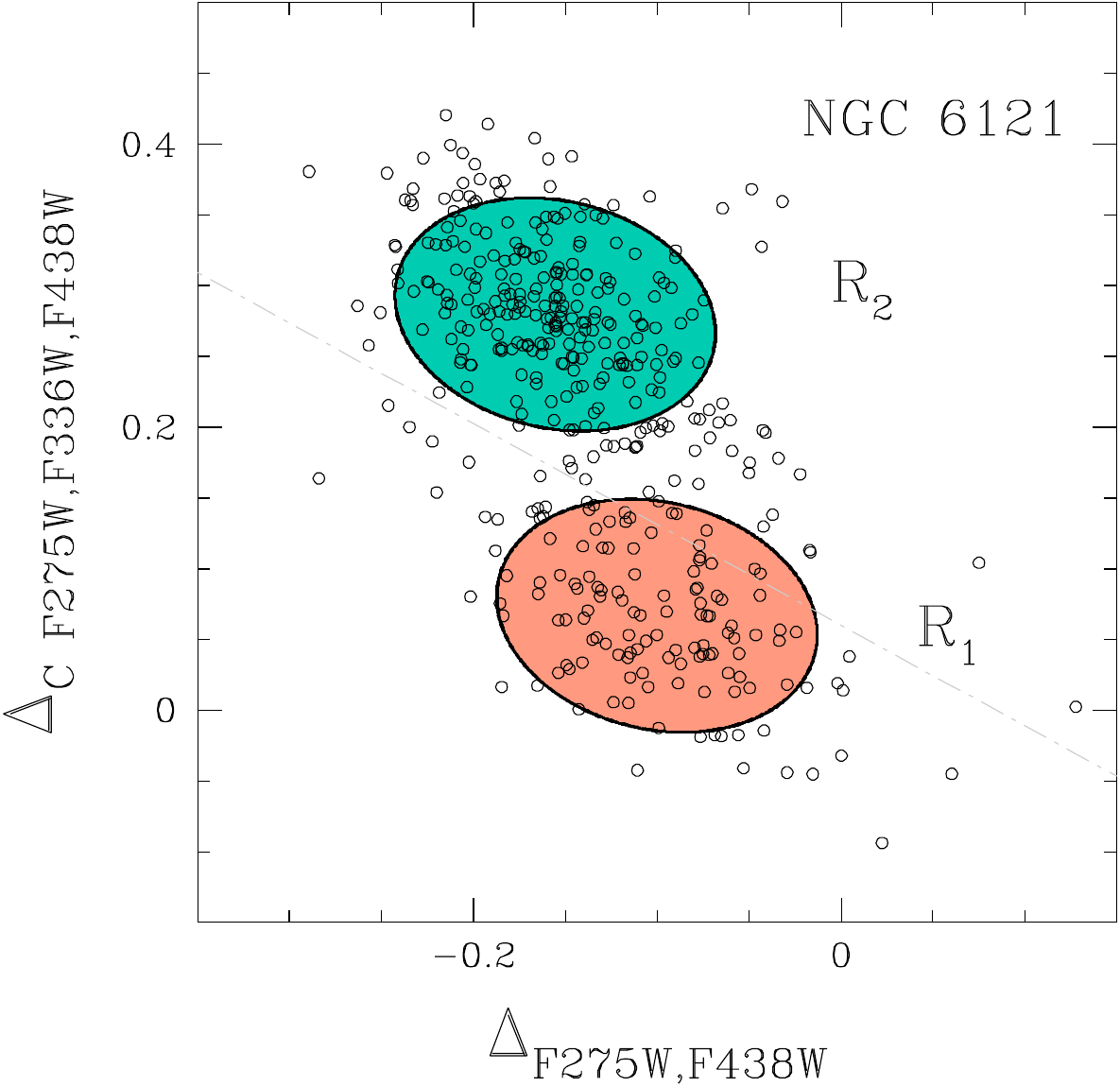}
  \caption{Reproduction of the ChM of M\,4 of Figure 5. The $R_{\rm 1}$ and $R_{\rm 2}$ regions are colored in red and blue, respectively (see text for details).}
 \label{fig:chm} 
 \end{center} 
\end{figure}

To compute the LF we applied a procedure similar to the one adopted for NGC\,2808 on the ChM \citep[see also][]{zennaro2019a}. Briefly, we first identified a sample of bona-fide 1G and 2G stars as the ones with $\Delta_{\rm C F275W,F336W,F438W}$ smaller and bigger than 0.155, respectively. Then, we computed the median of the ChM coordinates of these two groups of stars and used them as centres of elliptical regions with axis equal to photometric uncertainties. These regions are represented in Figure~\ref{fig:chm}, in which $R_{\rm 1}$ and $R_{\rm 2}$ contain the bulk of 1G and 2G stars, respectively.

We divided the studied magnitude interval in 0.6-wide $m_{\rm F160W}$ bins, and adapted the set of equations \ref{eq} to a diagram with two regions only.

\begin{equation}\label{eq3}
    \resizebox{.45\hsize}{!}{$N_{\rm i} = N_{\rm 1G}f_{\rm i}^{\rm 1G} + N_{\rm 2G}f_{\rm i}^{\rm 2G}$},
\end{equation}
Here, $N_{\rm i}$ is the number of stars in a region $R_{\rm i}$, $N_{\rm 1G}$ and $N_{\rm 2G}$ the number of 1G and 2G stars, $f_{\rm i}^{\rm 1G}$ and $f_{\rm i}^{\rm 2G}$ the fraction of stars of the two populations that fall in $R_{\rm i}$.

To infer $f_{\rm i}^{\rm 1G}$ and $f_{\rm i}^{\rm 2G}$, we performed AS tests, simulating 40,000 1G stars and 40,000 2G stars and measuring the fractions of recovered AS that fall in each region. Finally, we solved the system of equations~\ref{eq3}, deriving $N_{\rm 1G}$ and $N_{\rm 2G}$. We find that their fractions are nearly constant, with average values of 0.35$\pm$0.03 and 0.65$\pm$0.03, respectively.

\subsection{Lower Main Sequence} \label{subsec:lMSM4}

To measure the LF of MPs in the lower MS of M\,4 we used the $m_{\rm F160W}$ vs\,$m_{\rm F110W} - m_{\rm F160W}$ CMD, following the same procedure adopted in Section~\ref{subsec:lowMSn2808}. As shown in Figure~\ref{fig:NIRcmd}, two populations of very low-mass stars are clearly distinguishable along the MS.
The LF has been computed for stars with $16.6 < m_{\rm F160W} < 20.0$, which is the luminosity interval where the bulk of 1G stars are well separated from the 2G sequence. 

\begin{figure}[htp]
\begin{center} 
  \includegraphics[height=12cm,clip]{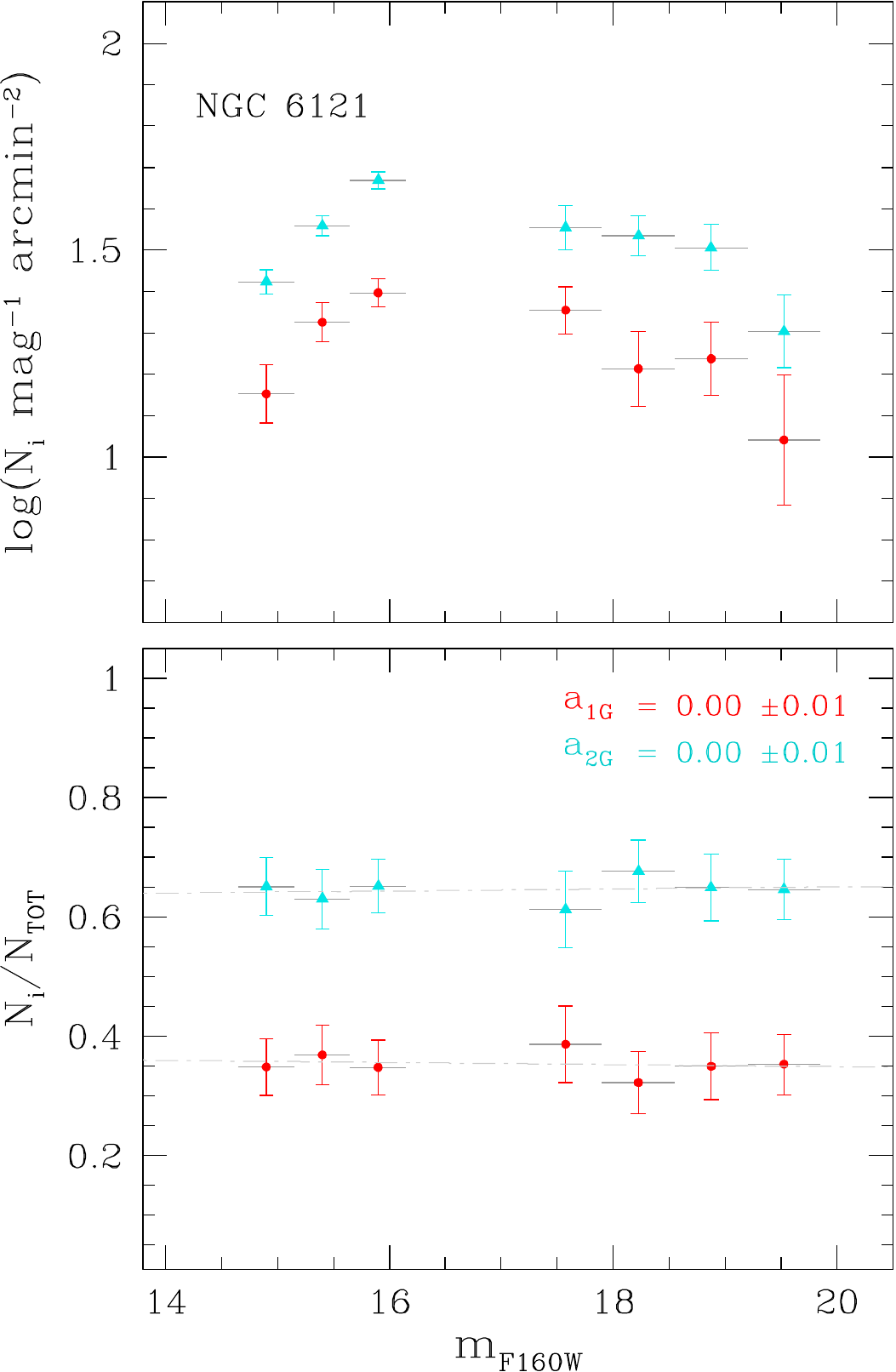}
  \caption{Same as Figure~\ref{fig:lf2808} but for M\,4. \\}
 \label{fig:lf6121} 
 \end{center} 
\end{figure}

Figure~\ref{fig:lf6121} shows the LFs of 1G and 2G stars, where the number of stars are normalized by unit of area and magnitude bin. The LF of both populations increases when moving from $m_{\rm F160W} \sim 14.5$ to $\sim 16.0$, and then decreases from $m_{\rm F160W} \sim 17.8$ to $m_{\rm F160W} \sim 19.8$, contrarily of what has been observed in NGC\,2808. This behaviour is shared by both 1G and 2G stars. The population ratios are in agreement with what has been observed in NGC\,2808, showing no variations with respect to the magnitude, confirmed by a {\it{p}}-value of 0.98 and a slope consistent with zero.

The average fraction of 1G and 2G populations in all the analyzed stars are 0.36$\pm$0.02 and 0.64$\pm$0.02, respectively.

\subsection{Mass Functions of multiple populations in \texorpdfstring{M\,4}{}} \label{sec:mf}

\begin{figure} 
\begin{center} 
  \includegraphics[height=12cm,clip]{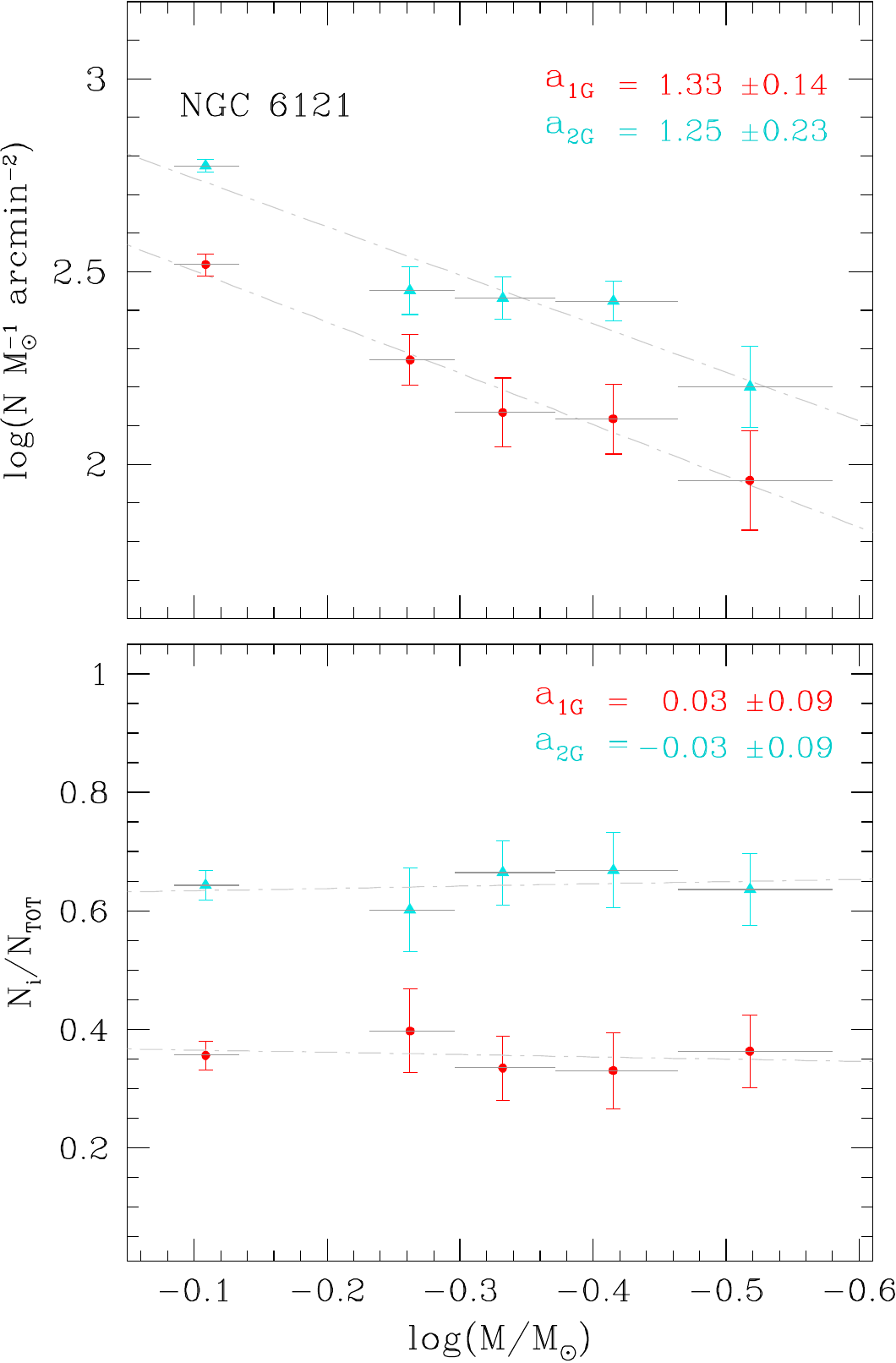}
  \caption{Same as Figure~\ref{fig:mf2808} but for 1G and 2G stars of M\,4.}
 \label{fig:mf6121} 
 \end{center} 
\end{figure}

To convert the LF of M\,4 stellar populations into MFs, we used isochrones with age 12.50 Gyr \citep[][]{dotter2010a}, [Fe/H] = $-$1.16 \citep[][2010 version]{harris1996a} and [$\alpha$/Fe] = 0.4 \citep[][]{dotter2010a}. Since the difference in helium mass fraction between 2G and 1G is $\Delta Y \sim 0.01$ \citep[e.g.][]{tailo2019}, and such small helium variation does not significantly affect the mass-luminosity relation, we assumed for all stars Y=0.246.
Results are plotted in the upper panel of Figure~\ref{fig:mf6121} and show that the MF of both populations share similar slopes. Moreover, as shown in the lower panel of  Figure~\ref{fig:mf6121}, the fractions of 1G and 2G stars are constant in the analyzed stellar mass interval, as confirmed by the high p-value of 0.97 and the slope values of the ratios.
 
As we did with for NGC2808, we repeated the MF measurement by considering equal-number bins to test if our results are affected by the differences between number of stars per bin. We obtained slope values of 1.24$\pm$0.29 and 1.11$\pm$0.39 for 1G and 2G stars, respectively, which are consistent with results obtained by using equal-width bins.

Although, this work is focused on the MPs within each GC, it is worth noticing that the MFs of stellar populations in NGC\,2808 and \,M4 follow different behaviours. Specifically, in NGC\,2808 they increase towards the lower masses, whereas M\,4 exhibits the opposite trend. 
    These behaviours are qualitatively consistent with the expected radial variation of the slope of the MF with the distance from the cluster's centre and the different regions covered by the analyzed fields in NGC\,2808 and M\,4. While for NGC\,2808 our FoVs  are located in the outer regions (at $\sim 2R_{\rm hm}$ where $R_{\rm hm}$ is the cluster 3D half- mass radius) from the GC centre, for M\,4 we studied stars closer to the cluster's centre (at $\sim 0.5 R_{\rm hm}$). As a cluster evolves, the effects of two-body relaxation \citep[e.g.,][]{spitzer1987} drive the segregation of massive stars towards the central regions and the migration of low-mass stars towards the outer regions. The 'inverted' slope of the MF in the central regions of M\,4 corresponds to a MF depleted in low-mass stars and is the manifestation of the effects of mass segregation.
    
    In both clusters we find only small and not statistically significant differences between the slope of the MF of 1G and 2G stars. As shown in the simulations of  \citet[][]{vesperini2018}, small differences in the slope of the 1G and 2G populations may arise during a cluster's evolution as a result of the differences in the initial structural properties of 1G and 2G stars when they form with the same IMF.  Stronger differences in the present-day MF are expected only when the 1G and the 2G population are characterized by large differences in their the IMFs. 
    Our observational results are thus consistent with those expected in systems in which the 1G and the 2G formed with similar IMFs.

\section{Radial distribution of multiple populations} \label{sec:rad}

\begin{table*}
    \caption{Fraction of MS-II stars in NGC\,2808 and fraction of 2G stars in M\,4. $R_{\rm min}$ and $R_{\rm max}$ are the minimum and maximum radial distances from the GC centre (in arcmin) of the stars used for each population-ratio measurement.}
    \footnotesize
    \centering
    \begin{tabular}{ccccccccccccccccc}
    \hline
    \hline
              & $R_{\rm min}$ & $R_{\rm max}$ & $N_{\rm MSII}/N_{\rm TOT}$ & & $R_{\rm min}$ & $R_{\rm max}$ & $N_{\rm MSII}/N_{\rm TOT}$ & & & & $R_{\rm min}$ & $R_{\rm max}$ & $N_{\rm 2G}/N_{\rm TOT}$\\
              \hline 
              & & & & & & & & & & \\
    NGC\,2808 & 0.00 & 0.60 & 0.55$\pm$0.03$^{a}$  & & 1.63 & 2.84 & 0.45$\pm$0.04$^{a}$ & & & NGC\,6121 & 0.00 &            1.69 & 0.71$\pm$0.01$^{c}$  \\
              & 0.60 & 0.82 & 0.55$\pm$0.02$^{a}$  & & 2.86 & 5.49 & 0.42$\pm$0.03$^{a}$ & & &           & 0.00 & 1.69 & 0.71$\pm$0.04$^{d}$  \\
              & 0.82 & 1.03 & 0.50$\pm$0.02$^{a}$  & & 4.20 & 6.37 & 0.33$\pm$0.04$^{b}$ & & &           & 0.63 & 3.31 & 0.64$\pm$0.02$^{b}$  \\
              & 1.03 & 1.63 & 0.52$\pm$0.02$^{a}$  & & 5.50 & 8.70 & 0.38$\pm$0.05$^{a}$ & & &         & 5.12 & 9.63 & 0.60$\pm$0.13$^{e}$  \\
              & & &   & & & & & &           & & 9.63 & 17.81 & 0.64$\pm$0.06$^{e}$ \\
    \hline
    \hline
    \\
    \end{tabular}
    \label{tab:rad}
    \footnotesize{
    
    References: $^{a}$\citet[][]{simioni2016a}; $^{b}$this work; $^{c}$\citet[][]{milone2020b}; $^{d}$\citet[][]{milone2017b}; $^{e}$\citet[][]{nardiello2015a}. }
\end{table*}

We now use the average population ratios inferred from this work to investigate the radial distribution of MPs in NGC\,2808 and M\,4. To do this, we combined our results with literature findings at different radial distances from the cluster centre.

\begin{figure*}
    \centering
    \includegraphics[width=16cm,clip]{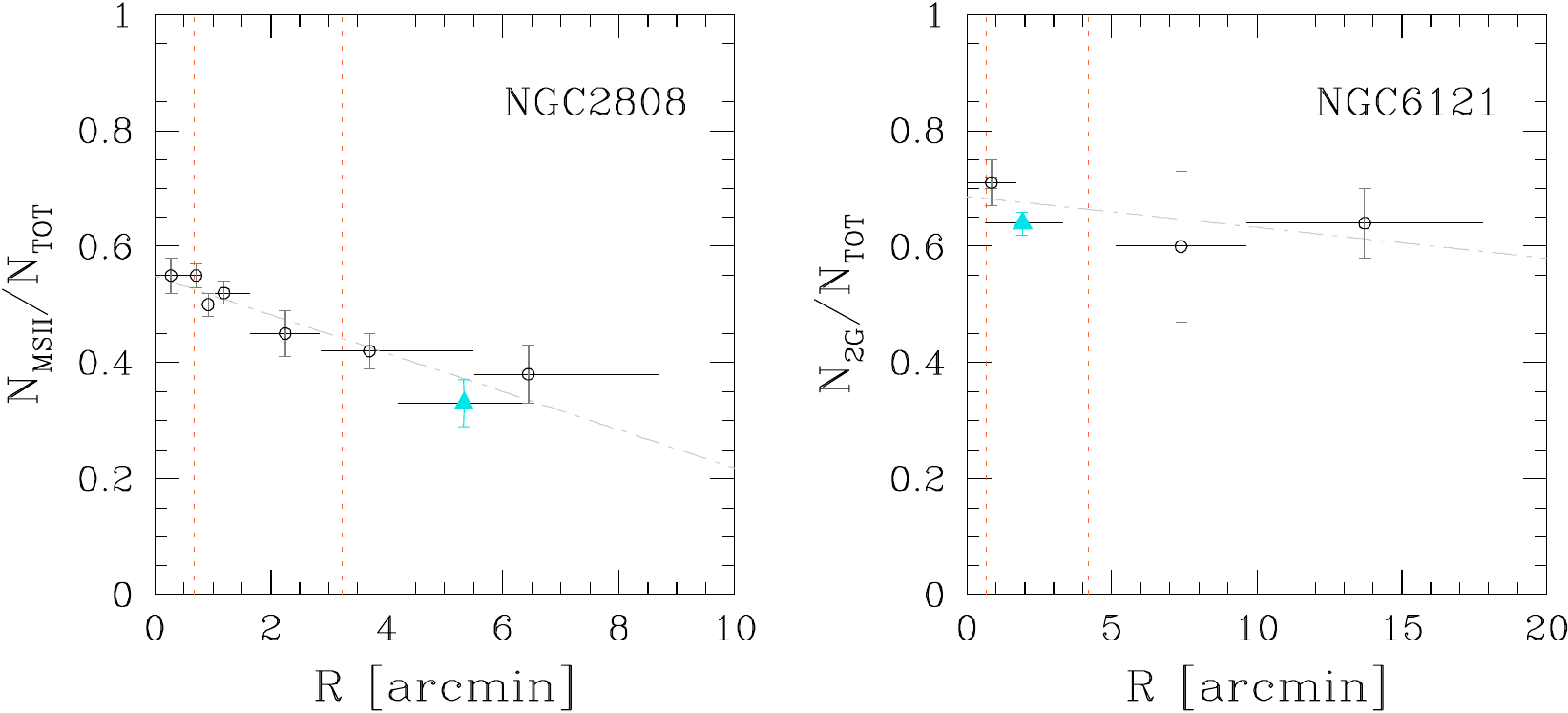}
    \caption{Radial distribution of the fraction of MS-II (left panel) and 2G stars (right panel) in NGC\,2808 and M\,4, respectively. Black circles represent literature results, while the cyan filled triangles show ratios inferred from this work. Black horizontal bars highlight the radial range covered by each measurements. The two dotted-vertical lines indicate the core and half-mass radius. We then show the best-fit straight lines (dash-dotted) in gray.}
    \label{fig:rad}
\end{figure*}

In Figure~\ref{fig:rad} we show the radial distribution of the fraction of the most extreme populations (MS-II and 2G stars for NGC\,2808 and M\,4, respectively). The azure filled triangles represent the average ratios inferred from this work, while the black circles display findings from literature. Details about the plotted data are listed in Table~\ref{tab:rad}. The population ratio estimates available for NGC\,2808, 
 cover radial distances to GC centre up to 8.70 arcmin ($\sim$2.7 $r_{\rm hm}$), while they reach 17.8 arcmin ($\sim$4.2 $r_{\rm hm}$) for M\,4.

The left panel of Figure\,\ref{fig:rad} displays the radial trend of the MS-II star fraction in NGC\,2808. We observe a clear decreasing radial trend, from $\sim$0.55 inside the core radius ($R_{\rm c}$) to $\sim$0.35 at $\sim$2.5$R_{\rm hm}$. This radial behaviour is confirmed by the slope of the best-fit straight line of $N_{\rm MSII}/N_{\rm TOT}$, -0.033$\pm$0.005.

The radial distribution of the fraction of 2G stars in M\,4 is plotted in the right panel of Figure\,\ref{fig:rad} and is consistent with a flat population ratio, as demonstrated by the slope of the best-fit straight line (-0.005$\pm$0.004). A possible exception is provided by the innermost bin, where the fraction of 2G stars is slightly higher than what is observed outside the core radius (0.71 and 0.63, respectively), but such difference is significant at $\sim 1 \sigma$-level only.

\section{Summary and conclusions} \label{sec:res}
The early study by \citet[][]{milone2012b} proved that the F110W and F160W filters of NIR/WFC3 on board {\it HST} are efficient tools to identify MPs of VLM stars.
The reason is that the F160W band is heavily affected by absorption from various molecules that contain oxygen (especially water), while F110W photometry is poorly affected by them. As a consequence, 2G stars, which are depleted in oxygen with respect to the 1G stars, have brighter F160W magnitudes and redder F110W$-$F160W colors. This finding has allowed us to extend the investigation of MPs among VLM stars as discussed in early works on NGC\,2808, M\,4, $\omega$\,Cen and NGC\,6752 \citep[e.g.][]{milone2012a, milone2014, milone2017a, milone2019a, dotter2015a}.

Here, we derived high-precision F110W and F160W photometry of nine Galactic GCs, namely NGC\,104 (47\,Tucanae), NGC\,288, NGC\,1851, NGC\,2808, $\omega$\,Cen, NGC\,5904 (M\,5), NGC\,6121 (M\,4), NGC\,6656 (M\,22) and NGC\,6752, and of the Galactic open cluster NGC\,6791, from images collected with the NIR/WFC3 camera on board {\it HST} and analyzed homogeneously. 
The resulting deep NIR CMDs allowed us to identify MPs below the MS knee in all GCs, thus providing 
 the first census of this phenomenon among M-dwarfs for a moderate sample of nine clusters. 
 
Stars fainter than the MS knee of all GCs exhibit intrinsic broadening in the $m_{\rm F110W}-m_{\rm F160W}$ color, which is associated with stellar populations with different oxygen abundances. We conclude that MPs are a widespread phenomenon among M-dwarfs of GCs, as observed for bright-MS and evolved stars \citep[i.e., SGB, RGB, HB and AGB stars, e.g.][]{carretta2009a, piotto2015a, milone2017a,  marino2019a, lagioia2021, dondoglio2021}.  Moreover, the presence of MPs among the M-dwarfs, which are fully-convective stars, corroborates the evidence that the chemical differences between 1G and 2G stars were present at their formation, as opposed to being established later.

The properties of MPs among M-dwarfs significantly change from one GCs to another.  The F110W$-$F160W MS width, $W_{\rm F110W,F160W}$, calculated 2.0 mag below the MS knee, ranges from $\sim$0.06 mag in M\,4 to $\sim$0.15 mag in $\omega$\,Centauri and correlates with the mass of the host cluster, thus corroborating similar conclusion inferred from the  RGB width \citep[e.g.][]{renzini2015,lagioia2019a, milone2017a, milone2020a} and from integrated light of GCs \citep[][]{jang2021a}. However, while the color broadening of M-dwarfs is mostly due to star-to-star oxygen variations, the F275W$-$F814W color and the $C_{\rm F275W,F336W,F814W}$ and $C_{\rm F275W,F336W,F814W}$ pseudo color used to analyze RGB stars are mainly sensitive to helium and nitrogen. Hence, our results suggest that dependence from cluster mass is a common property of helium, nitrogen and oxygen variations in GCs.  A remarkable difference is that the F110W$-$F160W broadening of M-dwarfs does not correlate with cluster metallicity, in contrast with the RGB width, which strongly correlates with [Fe/H]. We then found that $W_{\rm F110W,F160W}$ is deeply linked with the difference between the average [O/Fe] of 1G and 2G stars inferred by \citet[][]{marino2019a} from high-resolution spectroscopy of RGB stars, proving that oxygen variations in RGB and VLM stars are caused by the same phenomenon.
 
The F110W$-$F160W color distribution also varies from one cluster to another. NGC\,288, NGC\,2808, and M\,4 exhibit bimodal color distributions whereas in NGC\,6752 we distinguish a triple MS of VLM stars. 
These color distributions are qualitatively similar to the bimodal [O/Fe] distribution inferred from high-resolution spectroscopy of RGB stars of NGC\,288 and M\,4 \citep[][]{ carretta2009a, marino2008a, marino2011a} and the presence of three groups of stars with different [O/Fe] detected in NGC\,6752 \citep[][]{yong2005a, yong2015a}.
NGC\,2808 exhibits a much more complex MP pattern with at least five stellar populations with different oxygen abundances \citep[e.g.][]{milone2015, carretta2015a, marino2019a}. In this case, each group of MS-I and MS-II stars identified in the NIR CMD is composed of more than one stellar population that our data do not allow us to distinguish.
The majority of M-dwarfs in the $m_{\rm F160W}$ vs.\,$m_{\rm F110W}-m_{\rm F160W}$ CMD of $\omega$\,Centauri define a blue MS, but additional MS stars populate a broadened and red sequence. In the remaining GCs we notice a continuous color distribution with a predominance of M-dwarfs with blue colors. NGC\,6656, which seems to exhibit flatter color distribution is a possible exception. This seems consistent with results from spectroscopic investigations, where although these clusters exhibit internal oxygen variation, it is not possible to disentangle discrete populations. 
More-accurate photometry and/or spectroscopic oxygen determinations are mandatory to establish whether continue oxygen distribution is an intrinsic property of these clusters or is due to observational uncertainties \citep[e.g.][]{carretta2009a, marino2011b, marino2011c, johnson2010a, gratton2012a, cordero2014a}.
  
In contrast with what is observed in GCs, VLM stars of the open cluster NGC\,6791 distribute along a narrow sequence, where the F110W$-$F160W color spread is consistent with the broadening due to observational uncertainties. This result corroborates the idea that NGC\,6791 is a simple-population cluster, as suggested by previous studies based on high-resolution spectroscopy  \citep[e.g.,][]{bragaglia2014a, boberg2016a}.

In this paper, we investigate the LFs and MFs of MPs in NGC\,2808 and M\,4, which are the GCs where two distinct sequences of M-dwarfs are clearly visible below the MS knee. 

In the case of NGC\,2808 we investigated two groups of helium-poor and helium-rich stars that we named MS-I and MS-II stars, and correspond to the populations A$+$B$+$C and the populations D$+$E stars, identified by \citet[]{milone2015} in the cluster center.  

We combined NIR photometry, which is an efficient tool to disentangle MPs below the MS knee, and photometry in the F390W and F814W bands to identify and characterize the phenomenon along the upper MS. We analyzed three fields about 2 half-mass radii far away from the cluster center.
The main findings on NGC\,2808 can be summarized as follows:

\begin{itemize}
    
\item The fraction of MS-I and MS-II stars in NGC\,2808 are 0.67$\pm$0.04 and 0.33$\pm$0.04, respectively. When compared with the corresponding population ratios inferred from literature work based on {\it HST} data, we find that the fraction of MS-II stars increases  towards the cluster center, thus confirming the findings by \citet[][]{simioni2016a} that helium-rich stars of NGC\,2808 are more centrally concentrated that the helium-poor populations. 
    
\item We measured the LFs of MS-I and MS-II stars and inferred the corresponding MFs by using appropriate mass-luminosity relations from \citet[][]{dotter2008}, which account for the different helium abundances of the stellar populations. The LFs and MFs have been derived over the mass interval between $\sim$0.25 and $\sim$0.75 $\mathcal{M}_{\odot}$.
We find that the MFs of MS-I and MS-II exhibit similar shapes and the straight lines that provide the best fits of the MFs share the same slope.
Moreover, the fraction of MS-I and MS-II stars is constant in the whole analyzed interval of mass and luminosity. 

\end{itemize}

The analyzed FoV on M\,4 is located around 0.5 half-mass radii from the cluster center. We identified distinct sequences of 1G and 2G stars below the MS knee, in the interval between $\sim 0.3$ and $\sim$0.55 $\mathcal{M}_{\odot}$ by means of F110W and F160W NIR/WFC3 photometry. Moreover, we exploited the ChM to detect 1G and 2G stars along the upper MS, in the $\sim 0.7-0.8 \mathcal{M}_{\odot}$ mass range.
The most-relevant results on MPs in M\,4 include:

\begin{itemize}
    \item The fractions of 1G and 2G stars are 0.36$\pm$0.02 and 0.64$\pm$0.02,  respectively and are consistent with literature values inferred at similar distances from cluster centre. No significant radial variation is found in the fractions of 1G and 2G stars suggesting that the two populations are completely mixed in this clusters (with the possible exception of the innermost regions where there might be a slight enhancement in the fraction of 2G stars but the difference we found is only at 1-$\sigma$ level).
    
    \item We derived the LFs and MFs of 1G and 2G stars and find that they follow similar trends. In particular, the MFs of 1G and 2G stars have similar slopes and the population ratios is constant over the entire analyzed luminosity and mass intervals.
\end{itemize}

Our results are in conflict with scenarios in which the 2G would form from Bondi-Hoyle accretion on protostellar objects; in that case one might expect a strong difference between the 2G IMF (slope $\sim -2$)  and a 1G forming, for example, with a standard Kroupa (2001) IMF (slope $-1.3$) in the low-mass regime. This is also corroborated by the evidence that the relative numbers of 1G and 2G stars do not depend on stellar luminosity and mass. 
Our findings on NGC\,2808 and M\,4, together with previous results that the fractions of stars and the chemical composition of the three stellar populations of NGC\,6752 does not change with stellar mass \citep[][]{milone2019a}, indicate that the properties of MPs do not depend on stellar mass.
    
\citet[][]{vesperini2018} investigated the long-term evolution of the MF of MPs in GCs by means of N-body simulations.
Their simulations showed that small differences in the local and global present-day MF may arise as a result of the effects of dynamical evolution in systems where the 2G and the 1G stars formed with the same IMFs but in which the former were initially more centrally concentrated than the 1G stars. In the advanced stages of the evolution when the two populations are mixed these dynamically-induced differences will eventually vanish. Larger differences in the MF (particularly for dynamically older clusters) are instead found only in clusters in which the 1G and the 2G formed with significant differences in their IMFs.  In this context, the evidence from this paper that the distinct stellar populations of NGC\,2808 and M\,4 share similar MFs are consistent with a scenario where these populations originated with similar IMFs (at least in the low-mass range explored here).
If the 2G formed in a more dense and compact environment than 1G stars, as suggested by various scenarios for the formation of multiple populations \citep[e.g.,][]{dercole2010a, calura2019a}, the fact that 1G and 2G stars share the same MFs suggests that the low-mass star formation is not significantly affected by the density in the formation environment.

\acknowledgments
This work has received funding from the European Research Council (ERC) under the European Union's Horizon 2020 research innovation programme (Grant Agreement ERC-StG 2016, No 716082 'GALFOR', PI: Milone, http://progetti.dfa.unipd.it/GALFOR), and the European Union's Horizon 2020 research and innovation programme under the Marie Sklodowska-Curie (Grant Agreement No 797100). APM acknowledges support from MIUR through the FARE project R164RM93XW SEMPLICE (PI: Milone). Based on observations with the NASA/ESA \textit{Hubble Space Telescope}, obtained at the Space Telescope Science Institute, which is operated by AURA, Inc., under NASA contract NAS 5-26555. A.B. and EV acknowledge support from STScI grant GO-16289. EV acknowledges support from NSF grant AST-2009193.

\bibliography{sample63}{}
\bibliographystyle{aasjournal}

\end{document}